\documentclass[a4paper]{article}

\usepackage{INTERSPEECH2020}
\usepackage{subcaption}
\usepackage{color,cite}
\usepackage{hyperref}
\usepackage{makecell}
\usepackage{balance}
\usepackage{amsmath}
\usepackage{xcolor,colortbl}

\hypersetup{
   unicode=false,          
   pdftoolbar=true,        
   pdfmenubar=true,        
   pdffitwindow=false,     
   pdfstartview={FitH},    
   pdftitle={My title},    
   pdfauthor={Author},     
   pdfsubject={Subject},   
   pdfcreator={Creator},   
   pdfproducer={Producer}, 
   pdfkeywords={keyword1} {key2} {key3}, 
   pdfnewwindow=true,      
   colorlinks=true,       
   linkcolor=blue,          
   citecolor=blue,        
   filecolor=magenta,      
   urlcolor=blue           
}

\title{Predictions of Subjective Ratings and Spoofing Assessments of \\ Voice Conversion Challenge 2020 Submissions}
\name{Rohan Kumar Das$^1$, Tomi Kinnunen$^2$, Wen-Chin Huang$^3$, Zhenhua Ling$^4$, \\ 
    Junichi Yamagishi$^5$, Yi Zhao$^5$, Xiaohai Tian$^1$, Tomoki Toda$^3$}
\address{
  $^1$National University of Singapore, Singapore
  $^2$University of Eastern Finland, Finland \\
  $^3$Nagoya University, Japan
  $^4$University of Science and Technology of China, China \\
  $^5$National Institute of Informatics, Japan}
\email{vcc2020@vc-challenge.org}

\begin{document}
\maketitle
\begin{abstract}
The Voice Conversion Challenge 2020 is the third edition under its flagship that promotes intra-lingual semiparallel and cross-lingual voice conversion (VC). While the primary evaluation of the challenge submissions was done through crowd-sourced listening tests, we also performed an objective assessment of the submitted systems. The aim of the objective assessment is to provide complementary performance analysis that may be more beneficial than the time-consuming listening tests. In this study, we examined five types of objective assessments using automatic speaker verification (ASV), neural speaker embeddings, spoofing countermeasures, predicted mean opinion scores (MOS), and automatic speech recognition (ASR). Each of these objective measures assesses the VC output along different aspects. We observed that the correlations of these objective assessments with the subjective results were high for ASV, neural speaker embedding, and ASR, which makes them more influential for predicting subjective test results. In addition, we performed spoofing assessments on the submitted systems and identified some of the VC methods showing a potentially high security risk. 
\end{abstract}



\vspace{-1.5mm}
\section{Introduction}

\emph{Voice conversion} (VC), which refers to the digital cloning of a person's voice, can be used to modify an audio waveform so that it appears as if spoken by someone else (target) than the original speaker (source). VC is useful in many applications such as customizing audio book and avatar voices, dubbing, the movie industry, teleconferencing, singing voice modification, voice restoration after surgery, and the cloning of voices of historical persons. Since VC technology involves identity conversion, it can also be used to protect the privacy of individuals on social media and in sensitive interviews. For the same reason, VC also enables spoofing (fooling) voice biometric systems and therefore has potential security implications.  

VCC 2020 is the 3$^{\text{rd}}$ edition of the \emph{Voice Conversion Challenge} (VCC). While the general background and subjective results are provided in \cite{vcc2020summary}, this study focuses on complementary \emph{objective} evaluation results. 

Conventionally, the target of VC technology is human listeners, so subjective assessment has been the primary method of assessment in all the VCC challenges. On the other hand, progress has been made recently in research fields relevant to objective evaluation assessment, and human perception prediction and spoofing performance assessment against automatic speaker verification (ASV) systems are increasingly being used.  

The former is utilized to model and predict human perceptions automatically. Until recently, predicting human perception on synthetic speech has been challenging, but new research \cite{yoshimura,automos,Fu,8683175,Lo2019,Williams2020,choi2020deep,choi2020neural} has demonstrated that the advanced deep learning models and large amounts of paired data of synthetic speech and associated human judgement scores can lead to data-driven objective models that can predict human perception on synthetic speech to a certain extent. There are also several studies on using automatic speech recognition (ASR) as a proxy for subjective intelligibility estimation \cite{meyer2015autonomous}. 

The latter pertains to spoofing and anti-spoofing research for ASV. One goal of VC is to produce convincing mimicry of specific target speaker voices, and it is widely known that VC can fool (spoof) unprotected ASV systems \cite{spoof_review,bib:Attacker_overview2020}. Therefore, for increasing security and robustness, ASV systems normally adopt spoofing countermeasures, which are designed to learn the distinguishing artifacts present in spoofed audio produced by VC from human speech. We assume that spoofing performance against ASV may be correlated with the speaker similarity of VC systems and that the countermeasure (CM) performance represents the amount of artifacts produced by VC systems, which may or may not be audible to humans. Note that these systems are designed and optimized for discrimination by machines, and as such, the performances of ASV and CM may be different from human perceptions. 

With the above as our motivation, we provide an array of complementary objective results representative of recent objective evaluation techniques. Our tools include 
\begin{itemize}
    \item text-independent ASV \cite{Kaldi-xvector-recipe} for speaker similarity,
    \item text-independent CM \cite{spoof_review} for real-vs.-fake assessment,
    \item automatic MOS prediction \cite{Lo2019} for quality, and 
    \item ASR for intelligibility.
\end{itemize}
To the best of our knowledge, these four metrics have never before been examined as a group within any single VC study. Using these metrics, this paper investigates the two questions below: 
\begin{itemize}
    \item Can the metrics predict human judgements on naturalness and speaker similarity? 
    \item Which VC technology has the highest spoofing risk for ASV and CM? 
\end{itemize}

\begin{table*}[t]
    \centering
    \caption{Summary of evaluation metrics used in assessing VCC 2020 submissions. \textbf{ASV}: automatic speaker verification, \textbf{EER}: equal error rate, \textbf{MOS}: mean opinion score, \textbf{LCNN}: light convolutional neural network, \textbf{ASR}: automatic speech recognition, \textbf{WER}: word error rate, \textbf{CM}: countermeasure.}
     \resizebox{17.0cm}{!}{
    \begin{tabular}{|c|c|c|c|c|}
         \hline
         \textbf{Metric} & \textbf{Type of measure}            & \textbf{Measurement tool} & \textbf{Implementation} & \textbf{Metric interpretation}\\
         \hline\hline
         ASV EER & Conv. src $\leftrightarrow$ tgt similarity & ASV & Kaldi x-vector \cite{Kaldi-xvector-recipe} & Not similar = $0\%\dots 50\%$ = Similar \\
         $P_\text{fa}^\text{tar}$ & Conv. src $\leftrightarrow$ tgt similarity & ASV & Kaldi x-vector \cite{Kaldi-xvector-recipe} & Not similar = $0\%\dots100\%$ = Similar \\
         $P_\text{miss}^\text{src}$ & Conv. src $\leftrightarrow$ src similarity & ASV & Kaldi x-vector \cite{Kaldi-xvector-recipe} & Similar = $0\%\dots100\%$ = Not similar\\
         Cosine & Conv. src $\leftrightarrow$ tgt similarity & Speaker embedding & Kaldi x-vector \cite{Kaldi-xvector-recipe} & Not similar = $-1 \dots 1$ = Similar \\
         CM EER & Artifact assessment & Spoofing CM & LCNN~\cite{galinaITNTEERSPEECH2019} & Fake = $0\%\dots50\%$ = Real\\ 
         MOSNet & Quality & Objective MOS & MOSNet \cite{Lo2019} & Lowest = $1\dots5$ = Highest\\
         ASR WER & Intelligibility & ASR & Seq2seq with attention \cite{7472618}& Perfect = $0\%\dots100\%$ = Unintelligible\\
         \hline
    \end{tabular}
    }
    \label{tab:summary-of-objective-metrics}
\end{table*}

In Section 2 of this paper, we give an overview of the motivation behind each of the objective evaluation metrics. Their implementation details are described in Section 3. Correlations with human judgements are analyzed in Section 4, and spoofing performance against ASV and CM is discussed in Section 5. We conclude in Section 6 with a brief summary and mention of future work. 

\section{Methodology}

Objective metrics for speech signals can be categorized into \emph{intrusive} and \emph{non-intrusive} assessment methods. The former uses a ground-truth natural clean audio that has the same linguistic content as an input speech as a reference. The latter does not use any reference. A summary of the objective metrics used in this paper is provided in Table \ref{tab:summary-of-objective-metrics}. They are \emph{all} non-intrusive metrics. The following subsections provide the motivation and general description of each method. 
 
\subsection{Speaker Similarity: Automatic Speaker Verification}

We assess speaker similarity using ASV. An ASV system compares a test utterance of an unknown speaker with a hypothesized speaker's training utterance(s) and then outputs a speaker similarity score $s \in \mathbb{R}$. Higher scores indicate support for the \emph{same speaker} hypothesis and lower scores for the \emph{different speakers}. A hard decision is obtained by comparing $s$ with a pre-set verification threshold, $\tau^\text{asv}$. The speakers are declared to be the same if $s > \tau^\text{asv}$ (otherwise, different).

When VC methods achieve good mimicry of specific target speaker voices, the above score becomes higher and hence the converted audio will be judged as the same speaker. Therefore, we use the impact of VC on the ASV error rates as speaker similarity metrics here. For each VC system, we report three different kinds of ASV error: 
\begin{enumerate}
\item \textbf{Equal error rate} (EER): error rate at $\tau^\text{asv}$ at which the spoof false acceptance rate and target speaker miss rate equal each other;
\item \textbf{False acceptance rate of target} ($P_\text{fa}^\text{tar}$): proportion of converted utterances declared as the targeted speaker; 
\item \textbf{Miss rate of source} ($P_\text{miss}^\text{src}$): proportion of converted utterances \emph{not declared} as the original source speaker. 
\end{enumerate}

The first one gauges the ASV system's general accuracy in differentiating converted source utterances from real target speaker utterances (alternatively, the effectiveness of a VC in fooling ASV). The second metric also gauges the VC system's ability to fool the ASV system, with the difference that our false acceptance rate computation uses a threshold fixed prior to observing any VC samples. The same holds for the third metric, source miss rate, which gauges the VC system's ability to \emph{de-identify} the source speaker, in terms of ASV (alternatively, the ASV system's inability to \emph{re-identify} the source speaker). The reference values for ideal VC and useless VC are $(\text{EER}, P_\text{fa}^\text{tar},P_\text{miss}^\text{src})=(\frac{1}{2},1,1)$ and $(\text{EER},P_\text{fa}^\text{tar},P_\text{miss}^\text{src})=(0,0,0)$, respectively\footnote{Theoretically, EER is constrained between 0 and $\frac{1}{2}$ for any binary classification task, including ASV. Values larger than $\frac{1}{2}$ indicate decisions \emph{worse} than random guessing (label flip). In practice, the values may exceed $\frac{1}{2}$ due to data anomalies.}. 

The second and third error rates defined above are obtained by fixing $\tau^\text{asv}$ \emph{before} observing the converted utterances. The ASV system is optimized to differentiate real human same-speaker and different-speaker trials, but no VC samples are used in optimizing it. Thus, $\tau^\text{asv}$ is fixed using non-converted (natural) data only and remains fixed for the VC test. In practice, we set $\tau^\text{asv}$ at the EER operating point at which the (natural speech) miss and false alarm rates equal each other.

\subsection{Speaker Similarity: Neural Speaker Embedding}

Strictly speaking, the above ASV scoring procedures are different from how our subjects evaluated speaker similarity in the main listening test \cite{vcc2020summary}. In the listening test, they were asked to listen to a reference audio (which had different linguistic content from converted audio) and judge the speaker similarity of the converted audio to one of the reference audio files. 

Therefore, we computed the cosine similarity of speaker embedding vectors as well as the ASV scores. We extracted speaker embedding vectors from both converted audio and one of the reference audio files using the same tool as the above ASV system and measured their cosine similarity, $cos\_sim(A,B) = A \cdot B/\|A\|\|B\|$, where $A$ and $B$ are the speaker embedding vectors obtained from the converted audio and reference audio, respectively. Note that this is still technically a non-intrusive assessment because the reference audio for this measurement is different from the ground-truth natural clean audio that has the same linguistic content as an input speech.

\subsection{Artifact Assessment: Spoofing Countermeasures}

The spoofing countermeasures play an imperative role in defending against various attacks to ASV systems. In general, such countermeasures are designed to learn the distinguishing artifacts present in the natural human speech against different kinds of generated/spoofed speech to identify the spoofing attacks. Therefore, the artifact assessment using a spoofing countermeasure for the resultant speech of various VC systems could indicate the amount of artifacts of the converted speech. 

The performance of spoofing countermeasures is normally evaluated in terms of EER, similar to that of the ASV systems. However, unlike ASV, any human speaker speech is considered as target trials and the converted/spoofed speech serve as non-target trials. In the context of VC outputs, a high EER indicates the generation of more human-like speech, whereas a low EER indicates that the converted speech is inclined towards the characteristics of artificially generated speech.

\subsection{Quality: Objective Mean Opinion Scores}

The mean opinion score (MOS) used for subjective quality assessment is a numerical measure of the human-judged overall quality of audio, typically in the range of 1–5, where 1 is the lowest perceived quality and 5 is the highest. The objective MOS is used to assess speech quality by predicting and approximating the human assessment from an input audio. This technique has a long history and several metrics have been proposed for speech coding and telephony. Famous metrics include PESQ~\cite{rix2001perceptual} and POLQA~\cite{beerends2013perceptual}, both of which are intrusive assessments. There is also p536~\cite{malfait2006p}, a metric for non-intrusive assessment, but it is not designed for evaluating the quality of synthetic speech or converted speech.

When a large amount of paired data of synthetic speech and associated human judgement scores is available, we can view the objective MOS as a machine learning-based regression problem. Various deep learning models to predict the MOS values utilizing listening test data collected from the Blizzard Challenge \cite{king2014measuring}, our previous VCCs \cite{toda2016voice,Lorenzo-Trueba2018}, or the ASVspoof Challenge \cite{WANG2020101114} for supervision have been proposed. More specifically, \cite{yoshimura} conducted prediction of the Blizzard Challenge's results, \cite{Lo2019,choi2020deep,choi2020neural} reported predictions of VCC's results, and \cite{Williams2020} reported prediction results of ASVspoof's listening test results. While all of them exhibited moderate correlations with human judgments, it is still unknown whether these models can be generalized to new speech synthesis methods that are not already included in the databases.

Among these models, we have chosen to examine MOSNet~\cite{Lo2019}, which is a deep learning-based non-intrusive assessmentor, since it is reported that its prediction has a moderate correlation with the subjective evaluation done for the VCC 2018. 

\subsection{Intelligibility: Automatic Speech Recognition}
Although some of the recent VC methods can achieve high naturalness and similarity of converted speech, they may degrade the intelligibility of converted speech. For example, in the recognition-synthesis approach \cite{7900072, sun2016phonetic, miyoshi2017voice, ljliu2018wav} to VC, an ASR model is usually adopted to extract linguistic-related features, e.g., phonetic posteriorgrams (PPGs) \cite{sun2016phonetic} or bottleneck features \cite{ljliu2018wav}, from source speech. In this case, recognition errors are inevitable, which may degrade the intelligibility of the converted speech. Moreover, in the VC methods using sequence-to-sequence acoustic models \cite{8607053}, the failed alignment may lead to repetition and deletion of speech segments, especially when the amount of training data is limited \cite{zhang2019improving}.

Considering the cost of conducting subjective intelligibility evaluations for all conversion pairs, we adopted the word error rate (WER) of ASR as an objective metric on the intelligibility of converted speech in VCC 2020. A lower WER indicates a higher intelligibility. 

\section{Implementation Details}\label{sec:method-implementation}

\subsection{ASV and Cosine Distance of Speaker Embeddings}

Our ASV system utilizes x-vector \cite{Snyder2017-xvector}-based deep speaker embeddings. We use Kaldi's \cite{Povey2011-kaldi} recipe \cite{Kaldi-xvector-recipe} trained on VoxCeleb data \cite{Nagrani2020-voxceleb}. The system uses a time-delay neural network model (TDNN) trained with cross-entropy loss (treating training speakers as classes) to extract one 512-dimensional deep speaker embedding per utterance. The speaker similarity score is computed using probabilistic linear discriminant analysis (PLDA). 

The system is used as a scoring tool without specific modifications (e.g., domain adaptation) for the VCC 2020 data. The source and target speaker reference models are obtained from the respective training utterances provided to the challenge participants. The training x-vectors of each utterance are used to form one averaged x-vector per speaker. Test utterance x-vectors are then scored against these averaged models. For pre-VC ASV tests (required when setting the ASV threshold), we use the original source test data (provided to challenge participants) and the target speaker reference data (\emph{not} provided to challenge participants). For the VC tests, we replace the original source utterances with their VC-processed versions. 

For calculating the cosine distance of the speaker embeddings, we use the same Kaldi-based x-vector extractor as the ASV system. The x-vector dimensions were reduced to 200 using LDA before we compute the cosine similarity between converted speech and natural speech. We then calculate the averaged value per system.   

\subsection{Spoofing Countermeasure}

We used the light convolutional neural network (LCNN)-based system as a spoofing countermeasure in our studies~\cite{galinaITNTEERSPEECH2019}. The system considers 60-dimensional (20-static+20-$\Delta$+20-$\Delta\Delta$) linear frequency cepstral coefficient (LFCC) features as the input~\cite{ASVspoof2015_lfcc}. The training set of the ASVspoof 2019 logical access corpus is used to build the model~\cite{WANG2020101114}. The detailed architecture and implementation of the LCNN system is available in~\cite{LCNN_IS2020}. We considered the utterances from the training set of the VCC 2020 database as the bona fide trials, whereas the submissions from various teams constitute the spoof trials for evaluating the performance of every system submitted to the challenge. 

\subsection{MOSNet}

Our MOSNet model architecture followed the original setting in \cite{Lo2019}. Specifically, the raw magnitude spectrogram was first extracted from the converted speech and used as the input feature. The main model consisted of 12 convolution layers, one bidirectional long-short term memory layer, and two fully connected layers followed by a global averaging layer that pooled the frame-level scores to generate the final utterance level predicted MOS. The whole network was trained to regress the listening test scores by minimizing the mean square loss.

For training the MOSNet, we used two datasets: one composed of listening test data collected for the VCC 2018 \cite{Lorenzo-Trueba2018} and the other of listening test data collected for the ASVspoof 2019 \cite{WANG2020101114}. The former contains many of the VC systems available in 2018 and the latter contains more recent speech synthesis and VC methods available from 2019. They are referred to as MOSNet (vcc18)\footnote{\url{https://github.com/lochenchou/MOSNet}} and MOSNet (asvspoof19)\footnote{\url{https://github.com/rhoposit/MOS_Estimation2}}, respectively.

\subsection{ASR Engine}
The ASR engine was a prototype system developed by iFlytek. It features a state-of-the art end-to-end neural network-based ASR architecture and was trained using 10,000 hours of recordings and GB-level texts for language modeling. The vocabulary size was around 200,000. WERs were calculated by the \emph{HResults} tool in HTK using manual transcriptions as ground truth and considering substitutions, deletions, and insertions.

\begin{table*}[t!]
  \centering
  \caption{Performance of objective measures for Task 1 (intra-lingual semi-parallel VC). Red cells indicate top-5 systems (including ties) for each metric.}
    \label{tab:obj_Task1}
    \scriptsize
    \scalebox{0.97}{
    \begin{tabular}{|r|r|r|r|r|r|r|r|r|}
        \hline
\textbf{Team ID} & \textbf{ASV EER (\%)} & \textbf{ASV Pfa (\%)} & \textbf{ASV Pmiss (\%)} & \textbf{Cosine} & \textbf{CM EER (\%)} & \textbf{MOSNet (vcc18)} & \textbf{MOSNet (asvspoof19)} & \textbf{ASR WER (\%)}\\\hline\hline
T01 & 33.00 & 98.25 & 100.00 & 0.93 & 22.47 & 3.57 & 3.55 & 22.78\\
T02 & 14.00 & 87.50 & 100.00 & 0.86 & 26.74 & 3.32 & 3.22 & 12.32\\
T03 & 23.00 & 82.00 & 99.75 & 0.90 & 0.78 & 3.37 & 3.64 & 80.26\\
T04 & 45.13 & 99.00 & 100.00 & \cellcolor{red!25} 0.97 & 38.30 & \cellcolor{red!25} 3.92 & 3.25 & 22.84\\
T06 & 0.00 & 0.00 & 21.75 & 0.72 & 14.77 & 2.65 & 2.99 & \cellcolor{red!25} 3.65\\
T07 & \cellcolor{red!25} 48.50 & 99.75 & 100.00 & \cellcolor{red!25}0.96 & \cellcolor{red!25} 43.48 & 3.73 & 3.61 & 18.08\\
T08 & 0.50 & 0.50 & 78.25 & 0.76 & 37.97 & 2.89 & 2.86 & 6.95\\
T09 & 19.00 & 86.25 & 100.00 & 0.91 & 7.97 & 3.71 & 3.17 & 62.76\\
T10 & \cellcolor{red!25} 51.00 & \cellcolor{red!25} 100.00 & 100.00 & \cellcolor{red!25} 0.98 & \cellcolor{red!25} 43.98 & \cellcolor{red!25} 3.90 & \cellcolor{red!25} 3.70 & \cellcolor{red!25} 4.12\\
T11 & 38.50 & 99.00 & 99.50 & 0.94 & 42.75 & \cellcolor{red!25} 4.27 & \cellcolor{red!25} 4.17 & 5.43\\
T12 & 0.00 & 0.00 & 8.00 & 0.45 & 31.46 & 3.02 & 3.10 & \cellcolor{red!25} 3.50\\
T13 & 37.00 & 97.25 & 99.75 & 0.94 & 28.25 & 3.44 & 3.30 & 9.70\\
T14 & 1.00 & 6.00 & 99.50 & 0.76 & \cellcolor{red!25} 61.96 & 2.85 & 2.47 & 19.77\\
T16 & 33.00 & 97.00 & 100.00 & 0.93 & 36.51 & 3.33 & 3.33 & 21.52\\
T17 & 7.50 & 34.25 & 100.00 & 0.80 & \cellcolor{red!25} 47.24 & 2.95 & 3.11 & 52.07\\
T18 & 14.00 & 75.25 & 100.00 & 0.91 & 20.50 & 2.65 & 2.61 & 55.58\\
T19 & 33.63 & 98.50 & 100.00 & 0.94 & 42.75 & 3.08 & 3.39 & 65.80\\
T20 & 24.00 & 95.00 & 100.00 & 0.94 & 32.24 & 3.75 & 3.37 & 22.58\\
T21 & 2.00 & 13.75 & 98.50 & 0.75 & \cellcolor{red!25} 47.52 & \cellcolor{red!25} 3.86 & 3.45 & 30.84\\
T22 & \cellcolor{red!25} 52.00 & \cellcolor{red!25} 100.00 & 100.00 & \cellcolor{red!25}0.96 & 33.76 & 3.63 & 3.69 & 6.69\\
T23 & 45.00 & 99.75 & 100.00 & 0.94 & 32.02 & 3.51 & \cellcolor{red!25} 3.71 & 19.28\\
T24 & 25.50 & 98.50 & 100.00 & 0.91 & 20.50 & 3.54 & 3.51 & 23.83\\
T25 & 33.50 & 98.50 & 98.75 & \cellcolor{red!25}0.96 & 27.52 & 3.58 & 3.67 & 4.82\\
T26 & 3.63 & 53.00 & 99.25 & 0.86 & 18.98 & 3.76 & 3.38 & 28.04\\
T27 & 37.50 & \cellcolor{red!25} 100.00 & 100.00 & 0.95 & 19.49 & 3.41 & 3.31 & \cellcolor{red!25} 3.42\\
T28 & 34.50 & 96.00 & 99.75 & 0.95 & 32.70 & 3.58 & 3.25 & 96.17\\
T29 & \cellcolor{red!25} 45.50 & \cellcolor{red!25}100.00 & 100.00 & \cellcolor{red!25}0.96 & 34.44 & \cellcolor{red!25} 3.94 & \cellcolor{red!25} 3.71 & 8.47\\
T30 & \cellcolor{red!25} 46.00 & 99.75 & 100.00 & \cellcolor{red!25} 0.97 & 2.02 & 3.72 & 3.61 & \cellcolor{red!25} 2.77\\
T31 & 31.63 & 99.50 & 100.00 & 0.92 & 25.45 & 3.27 & 2.66 & 77.80\\
T32 & 18.00 & 95.00 & 100.00 & 0.94 & 30.55 & 3.48 & 3.55 & 4.21\\
T33 & 43.13 & \cellcolor{red!25} 100.00 & 100.00 & \cellcolor{red!25}0.96 & 33.25 & 3.55 & \cellcolor{red!25} 3.72 & 9.64\\       
\hline        
\end{tabular}}
\end{table*}

\section{Objective Evaluation Results for Each Submitted VC System}
\label{objEvalVCC2020}

We used converted audio produced by each of the submitted VC systems for Tasks 1 and 2 and assessed the objective evaluation metrics described earlier. It is noted that the Tasks 1 and 2 are intra-lingual semi-parallel and cross-lingual VC tasks, respectively. The results for each VC system for Tasks 1 and 2 are summarized in Tables \ref{tab:obj_Task1} and \ref{tab:obj_Task2}, respectively.

\begin{table*}[t!]
  \centering
  \caption{Performance of objective measures for Task 2 (cross-lingual VC). Red cells indicate top-5 systems (including ties) for each metric.}
    \label{tab:obj_Task2}
   \scriptsize
    \scalebox{0.97}{
    \begin{tabular}{|r|r|r|r|r|r|r|r|r|}
        \hline
\textbf{Team ID} & \textbf{ASV EER (\%)} & \textbf{ASV Pfa (\%)} & \textbf{ASV Pmiss (\%)} & \textbf{Cosine} & \textbf{CM EER (\%)} & \textbf{MOSNet (vcc18)} & \textbf{MOSNet (asvspoof19)} & \textbf{ASR WER (\%)}\\\hline\hline
T02 & 19.18 & 60.50 & 98.50 & 0.82 & 22.15 & 3.39 & 2.96 & 12.97\\
T03 & 16.00 & 43.50 & 99.83 & 0.84 & 0.82 & 3.31 & 3.67 & 81.25\\
T05 & 25.63 & 79.50 & 99.67 & 0.90 & 13.48 & 2.78 & 2.09 & 6.48\\
T06 & 1.18 & 1.33 & 21.33 & 0.73 & 16.01 & 2.80 & 2.98 & 5.18\\
T07 & \cellcolor{red!25} 60.37 & \cellcolor{red!25} 100.00 & 99.00 & 0.91 & \cellcolor{red!25} 44.49 & 3.68 & 3.55 & 24.82\\
T08 & 0.08 & 0.17 & 72.83 & 0.74 & \cellcolor{red!25} 46.64 & 3.00 & 3.07 & 3.80\\
T09 & 25.92 & 85.00 & 99.83 & 0.86 & 7.15 & 3.71 & 3.14 & 65.85\\
T10 & \cellcolor{red!25} 45.55 & 97.50 & 96.00 & \cellcolor{red!25} 0.95 & \cellcolor{red!25} 49.81 & 3.96 & \cellcolor{red!25} 3.72 & \cellcolor{red!25} 4.11\\
T11 & 41.55 & \cellcolor{red!25} 98.83 & 93.67 & 0.91 & 42.97 & \cellcolor{red!25} 4.26 & \cellcolor{red!25} 4.17 & 5.96\\
T12 & 26.00 & 71.33 & 100.00 & 0.84 & 29.81 & 2.81 & 2.31 & 29.40\\
T13 & 36.37 & 90.50 & 97.33 & 0.90 & 21.51 & 3.55 & 3.47 & 6.46\\
T15 & 4.82 & 17.00 & 98.00 & 0.86 & \cellcolor{red!25} 50.50 & \cellcolor{red!25} 4.33 & 3.30 & 13.10\\
T16 & 41.18 & 95.17 & 99.67 & 0.88 & 34.36 & 3.29 & 3.02 & 25.43\\
T18 & 20.37 & 66.00 & 99.67 & 0.84 & 32.02 & 2.75 & 2.27 & 74.01\\
T19 & \cellcolor{red!25} 44.00 & \cellcolor{red!25} 98.67 & 100.00 & 0.87 & 38.35 & 3.24 & 3.31 & 76.77\\
T20 & 5.63 & 18.67 & 91.00 & 0.85 & 34.68 & \cellcolor{red!25}4.06 & 3.61 & 23.15\\
T22 & 30.82 & 89.50 & 100.00 & 0.85 & 42.97 & 3.55 & 3.64 & 30.96\\
T23 & 32.82 & 88.83 & 97.50 & 0.91 & \cellcolor{red!25} 53.67 & 3.31 & 2.87 & 18.32\\
T24 & \cellcolor{red!25} 48.82 & \cellcolor{red!25} 99.33 & 99.33 & 0.88 & 17.97 & 3.83 & 3.53 & 45.11\\
T25 & 30.82 & 89.83 & 90.33 & \cellcolor{red!25} 0.93 & 29.30 & 3.60 & \cellcolor{red!25} 3.70 & \cellcolor{red!25} 4.58\\
T26 & 4.37 & 15.67 & 97.50 & 0.80 & 22.97 & \cellcolor{red!25} 4.14 & 3.36 & 34.58\\
T27 & 33.63 & 75.33 & 93.17 & 0.89 & 26.64 & 3.37 & 3.47 & \cellcolor{red!25} 3.93\\
T28 & 18.82 & 48.17 & 88.83 & 0.87 & 34.17 & 3.49 & 3.35 & 72.41\\
T29 & \cellcolor{red!25} 47.63 & \cellcolor{red!25} 98.83 & 98.83 & \cellcolor{red!25} 0.93 & 33.85 & \cellcolor{red!25} 3.98 & \cellcolor{red!25} 3.74 & 8.86\\
T30 & 40.00 & 92.17 & 96.33 & \cellcolor{red!25} 0.94 & 2.02 & 3.47 & \cellcolor{red!25} 3.70 & \cellcolor{red!25} 3.21\\
T31 & 29.63 & 90.83 & 99.17 & 0.86 & 19.81 & 3.21 & 2.90 & 70.02\\
T32 & 15.63 & 64.83 & 98.50 & \cellcolor{red!25} 0.92 & 28.98 & 3.54 & 3.44 & \cellcolor{red!25} 5.14\\
T33 & 23.63 & 80.33 & 80.67 & 0.89 & 34.49 & 3.92 & 3.53 & 19.55\\      
\hline        
\end{tabular}}
\end{table*}

\vspace{1ex}
\noindent
\textbf{ASV:} The ASV results are shown in the first to fourth columns in the tables. Concerning false acceptance rates on Task 1, 20 (out of 31) systems achieved error rates higher than 90\%. Further, the top-5 systems obtained perfect results (100\% false acceptance). Many other teams obtained near-perfect results. Concerning the miss rate of converted source speakers, it was higher than 90\% for all teams except three (06, 08, 12). To sum up, the top systems all achieved similar results, and the majority of the systems achieved a high target speaker similarity. Nearly all systems managed to successfully move the converted voice `away' from the original source. The results were more varied for Task 2, however: only ten systems yielded false acceptance rates above 90\%, and there was substantial variation across the systems. The miss rates were generally worse than those in Task 1 for most systems, although, similar to Task 1, for most systems they were reasonably high. 

\vspace{1ex}
\noindent\textbf{Spoofing Countermeasures:} 
The fifth column of the tables shows the performance of the LCNN-based spoofing countermeasure for all the submitted systems on both tasks of VCC 2020. We can see that most of the teams achieved a high EER, which indicates that the VC systems were able to generate natural human-like speech that was not easily detectable by the spoofing countermeasure. In addition, the performance trends of the spoofing countermeasure for most teams were similar in both tasks, excluding four teams (08, 18, 22 and 23). They also showed a relatively higher EER for Task 2 than Task 1, which might be a result of the very different settings used for both tasks by those teams. 

\vspace{1ex}
\noindent\textbf{MOSNet:}
The MOSNet predictions of all systems are shown in the sixth column in the tables. We can see that the MOSNet predictions fell between 2.5 to 4.5, while the ground truth MOS typically ranged from 1.0 to 4.5, indicating that the overall variance of the MOSNet predictions was rather small. We can also see that the scores of each team for Tasks 1 and 2 were similar. This is consistent with the fact that most teams utilized the same system for both tasks.    %

\noindent \textbf{WER:} The ASR WERs of all systems are shown in the final column in the tables. We can observe a large variance of WERs among teams in both tasks. For example, in Task 1, seven teams had WERs that were lower than 5\%, while seven other teams had WERs that were higher than 50\%. This indicates the diversity of the conversion methods adopted by different teams. After subjectively examining a few samples from the teams with high WERs, we can clearly perceive their intelligibility degradation. Comparing the WERs in Tasks 1 and 2, as expected, most teams achieved a higher WER on Task 2, which indicates the difficulty of cross-lingual VC.

\begin{table*}[t]
  \centering
  \caption{Pearson correlation coefficients with subjective evaluation results for each metric and ($p$-values). Top: correlation with subjective scores from \textbf{English listeners}. Bottom: correlation with subjective scores from \textbf{Japanese listeners}. Bold font indicates the highest correlation among the objective metrics.}
    \label{tab:correlation}
    \footnotesize
    \scalebox{0.96}{
    \begin{tabular}{|c|c c c c c c c|}
        \hline
        \textbf{\makecell{Subjective\\score (ENG)}} & \textbf{\makecell{ASV \\EER (\%)}} & \textbf{\makecell{ASV \\Pfa (\%)}} & \textbf{Cosine distance} & \textbf{\makecell{Countermeasure\\EER (\%)}} & \textbf{\makecell{MOSNet \\ (vcc18)}} & \textbf{\makecell{MOSNet\\(asvspoof19)}} & \textbf{\makecell{ASR\\WER (\%)}} \\
        \hline\hline
            Task 1 MOS & \textbf{0.70} ($p<$0.01) & 0.53 ($p<$0.01) & 0.42 ($p>$0.01) & 0.00 ($p>$0.01) & 0.52 ($p<$0.01)) & 0.66 ($p<$0.01) & $-$0.65 ($p<$0.01) \\
            Task 1 SIM & \textbf{0.89} ($p<$0.01) & 0.82 ($p<$0.01) & 0.85 ($p<$0.01) & 0.07 ($p>$0.01)  & 0.54 ($p<$0.01) & 0.61 ($p<$0.01) & $-$0.18 ($p>$0.01)\\
            Task 2 MOS &          0.34 ($p>$0.01) & 0.26 ($p>$0.01) & 0.59 ($p<$0.01) & 0.27 ($p>$0.01)  & 0.43 ($p>$0.01) & 0.58 ($p<$0.01) & \textbf{$-$0.73} ($p<$0.01) \\
            Task 2 SIM & \textbf{0.90} ($p<$0.01) & 0.86 ($p<$0.01) & 0.82 ($p<$0.01) & 0.19 ($p>$0.01)  & 0.23 ($p>$0.01) & 0.32 ($p>$0.01) & $-$0.14 ($p>$0.01)\\
        \hline
    \end{tabular}} \\
    \vspace{3mm}
    \scalebox{0.96}{
    \begin{tabular}{|c|c c c c c c c|}
        \hline
        \textbf{\makecell{Subjective\\score (JPN) }} & \textbf{\makecell{ASV \\EER (\%)}} & \textbf{\makecell{ASV \\Pfa (\%)}} & \textbf{Cosine distance} & \textbf{\makecell{Countermeasure\\EER (\%)}} & \textbf{\makecell{MOSNet \\ (vcc18)}} & \textbf{\makecell{MOSNet\\(asvspoof19)}} & \textbf{\makecell{ASR\\WER (\%)}} \\
        \hline\hline
            Task 1 MOS & \textbf{0.80} ($p<$0.01) & 0.64 ($p<$0.01) & 0.54 ($p<$0.01)         & $-$0.03 ($p>$0.01) & 0.54 ($p<$0.01) & 0.70 ($p<$0.01) & $-$0.45 ($p=$0.01)\\
            Task 1 SIM & \textbf{0.88} ($p<$0.01) & 0.82 ($p<$0.01) & 0.86 ($p<$0.01)         & 0.07 ($p>$0.01)  & 0.54 ($p<$0.01) & 0.62 ($p<$0.01) & $-$0.13 ($p>$0.01)\\
            Task 2 MOS &        0.40 ($p>$0.01)   & 0.31 ($p>$0.01) & 0.64 ($p<$0.01)         & 0.29 ($p>$0.01)  & 0.48 ($p<$0.01) & \textbf{0.67} ($p<$0.01) & $-$0.59 ($p<$0.01)\\
            Task 2 SIM & 0.83 ($p<$0.01)          & 0.82 ($p<$0.01) & \textbf{0.86} ($p<$0.01)& 0.29 ($p>$0.01)  & 0.24 ($p>$0.01) & 0.35 ($p>$0.01) & $-$0.26 ($p>$0.01)\\
        \hline
    \end{tabular}}
\end{table*}

\section{Can the Metrics Predict Human Judgements on VC Speech?}
\subsection{Correlation with subjective evaluation results}

In this section, we investigate our first question, ``Can the metrics predict human judgements on the naturalness and speaker similarity of converted audio submitted for VCC 2020?" 

In the VCC 2020, we conducted two large-scale crowd-sourced listening tests on the naturalness and speaker similarity of converted speech. The first test was done by 68 native English listeners (32 female, 33 male, and 3 unknown) and the second by 206 native Japanese listeners (96 male and 110 female). More details are described in \cite{vcc2020summary}. Using these listening test results, we measured the correlations of each objective evaluation metric with the subjective evaluation results by the English and Japanese listeners.

More specifically, we created scatter plots matching each of the objective metrics at the system level and each of the subjective evaluation results, then calculated the Pearson correlation coefficients. The scatter plots are shown in Appendix \ref{sub-vs-obj-scatter}. 

Table \ref{tab:correlation} shows the Pearson correlation coefficients with subjective evaluation results for each metric along with their $p$-values. The top and bottom tables show the correlations with the subjective scores obtained from the English and Japanese listeners, respectively. We first summarize the correlation analysis using the English listeners and then discuss the difference between this case and the Japanese one.

\vspace{1ex}
\noindent\textbf{Subjective quality rating:}
From Table 4, we can see that the ASV-related metrics (EER, Pfa), MOSNet (vcc18, asvspoof19), and ASR WER had moderately positive or negative correlations with the subjective quality ratings in Task 1, while cosine distance, MOSNet (asvspoof19), and ASR WER had moderate correlations in Task 2. These findings are statistically significant. While it is slightly surprising to see the ASV-related metrics and cosine distance were correlated with the subjective quality ratings, we assume this stems from the fact that human judgements on quality and speaker similarity are not independent. The fact that the ASV-related metrics had higher correlations with the speaker similarity ratings also underpins this. We can also see that MOSNet (asvspoof19) had higher correlations than MOSNet (vcc18). This was because the asvspoof19 dataset contains more diverse and new speech generation methods than the VCC18 dataset, which demonstrates the importance of choosing the appropriate training dataset for MOSNet.  

\noindent\textbf{Subjective speaker similarity rating:}
We can see that all of the ASV-related metrics (EER, Pfa, cosine distance) had strong correlations with subjective speaker similarity ratings in both tasks. Among these, the EER had the highest correlations ($r=0.89$ for Task 1 and $r=0.90$ for Task 2). MOSNet (vcc18, asvspoof19) had a moderate correlation with the speaker similarity ratings in Task 1, but its correlation in Task 2 was not statistically significant. 

\begin{table*}[t]
  \centering
  \caption{Breakdown per language of Pearson correlation coefficients with subjective evaluation results for each metric along with $p$-values. Top: correlation with subjective scores from \textbf{English listeners}. Bottom: correlation with subjective scores from \textbf{Japanese listeners}. Bold font indicates the highest correlation among the objective metrics. F, G, and M represent Finnish, German, and Mandarin target speakers, respectively.}
    \label{tab:correlation-breakdown}
    \footnotesize
    \scalebox{0.94}{
    \begin{tabular}{|c|c c c c c c c|}
        \hline
        \textbf{\makecell{Subjective\\score (ENG)}} & \textbf{\makecell{ASV \\EER (\%)}} & \textbf{\makecell{ASV \\Pfa (\%)}} & \textbf{Cosine distance} & \textbf{\makecell{Countermeasure\\EER (\%)}} & \textbf{\makecell{MOSNet \\ (vcc18)}} & \textbf{\makecell{MOSNet\\(asvspoof19)}} & \textbf{\makecell{ASR\\WER (\%)}} \\
        \hline\hline
            Task 2 MOS (F) & 0.26 ($p>$0.01) & 0.26 ($p>$0.01) & 0.62 ($p<$0.01) & 0.40 ($p>$0.01) & 0.42 ($p>$0.01) & 0.41 ($p>$0.01) & \textbf{$-$0.74} ($p<$0.01)\\
            Task 2 MOS (G) & 0.33 ($p>$0.01) & 0.27 ($p>$0.01) & 0.45 ($p>$0.01) & 0.16 ($p>$0.01) & 0.47 ($p>$0.01) & 0.47 ($p>$0.01) & \textbf{$-$0.71} ($p<$0.01)\\
            Task 2 MOS (M) & 0.36 ($p>$0.01) & 0.26 ($p>$0.01) & 0.64 ($p<$0.01) & 0.20 ($p>$0.01) & 0.36 ($p>$0.01) & 0.34 ($p>$0.01) & \textbf{$-$0.72} ($p<$0.01)\\
            Task 2 SIM (F) & \textbf{0.84} ($p<$0.01) & 0.81 ($p<$0.01) & 0.70 ($p<$0.01) & 0.27 ($p>$0.01) & 0.25 ($p>$0.01) & 0.26 ($p>$0.01) & $-$0.09 ($p>$0.01)\\
            Task 2 SIM (G) & 0.82 ($p<$0.01) & 0.82 ($p<$0.01) & \textbf{0.84} ($p<$0.01) & 0.08 ($p>$0.01) & 0.38 ($p>$0.01) & 0.37 ($p>$0.01) & $-$0.26 ($p>$0.01)\\
            Task 2 SIM (M) & \textbf{0.88} ($p<$0.01) & 0.86 ($p<$0.01) & 0.74 ($p<$0.01) & 0.09 ($p>$0.01) & 0.06 ($p>$0.01) & 0.06 ($p>$0.01) & $-$0.02 ($p>$0.01)\\
        \hline
    \end{tabular}} \\
    \vspace{3mm}
    \scalebox{0.94}{
    \begin{tabular}{|c|c c c c c c c|}
        \hline
        \textbf{\makecell{Subjective\\score (JPN) }} & \textbf{\makecell{ASV \\EER (\%)}} & \textbf{\makecell{ASV \\Pfa (\%)}} & \textbf{Cosine distance} & \textbf{\makecell{Countermeasure\\EER (\%)}} & \textbf{\makecell{MOSNet \\ (vcc18)}} & \textbf{\makecell{MOSNet\\(asvspoof19)}} & \textbf{\makecell{ASR\\WER (\%)}} \\
        \hline\hline
            Task 2 MOS (F) & 0.34 ($p>$0.01) & 0.34 ($p>$0.01) & \textbf{0.67} ($p<$0.01) &	0.42 ($p>$0.01) & 0.46 ($p>$0.01) & 0.46 ($p>$0.01) &	$-$0.62 ($p<$0.01)\\
            Task 2 MOS (G) & 0.41 ($p>$0.01) & 0.32 ($p>$0.01) & 0.51 ($p<$0.01) & 0.20 ($p>$0.01) & 0.50 ($p<$0.01) & 0.50 ($p>$0.01) & \textbf{$-$0.55} ($p<$0.01)\\
            Task 2 MOS (M) & 0.40 ($p>$0.01) & 0.25 ($p>$0.01) & \textbf{0.65} ($p<$0.01) &	0.20 ($p>$0.01) & 0.43 ($p>$0.01) & 0.42 ($p>$0.01) &	$-$0.56 ($p<$0.01)\\
            Task 2 SIM (F) & 0.75 ($p<$0.01) & 0.75 ($p<$0.01) & \textbf{0.81} ($p<$0.01) &	0.36 ($p>$0.01) & 0.27 ($p>$0.01) & 0.27 ($p>$0.01) &	$-$0.29 ($p>$0.01)\\
            Task 2 SIM G) & 0.82 ($p<$0.01) & 0.82 ($p<$0.01) & \textbf{0.85} ($p<$0.01) &	0.18 ($p>$0.01) & 0.34 ($p>$0.01) & 0.35 ($p>$0.01) &	$-$0.27 ($p>$0.01)\\
            Task 2 SIM (M) & \textbf{0.81} ($p<$0.01) & 0.77 ($p<$0.01) & 0.78 ($p<$0.01) &	0.25 ($p>$0.01) & 0.11 ($p>$0.01) & 0.11 ($p>$0.01) &	$-$0.19 ($p>$0.01)\\
        \hline
    \end{tabular}}
\end{table*}

\noindent\textbf{English listeners vs.\ Japanese listeners:}
Next, we analyzed the differences between the English listeners' case and the Japanese one. We can see that the general tendencies were the same in both cases; that is, ASV-related metrics had strong correlations with subjective speaker similarity ratings, and MOSNet and ASR WER had moderately negative correlations with subjective quality ratings. Two minor differences were that the cosine distance had a slightly higher correlation than ASV EER for Task 2 and that MOSNet (asvspoof19) had a higher correlation than ASR WER for Task 2. These differences were marginal.

\subsection{Analysis of objective speaker similarity scores of top-ranked VC submissions}

We have demonstrated that the ASV metrics, in particular EER, had a strong correlation with subjective ratings on speaker similarity. Therefore, it should prove fruitful to analyze the top-ranked VC submissions further, as their subjective speaker similarity in Task 1 was as good as the target speakers according to the listening test results \cite{vcc2020summary} and hence we cannot obtain meaningful differences between the top submissions from the listening test only. 

As reported in \cite{vcc2020summary}, eight VC systems (T10, T22, T27, T13, T33, T23, T29, and T07) had statistically fewer significant differences from human speech. As expected, some of these differences related to the EERs (as shown in Table \ref{tab:obj_Task1}), although all of them had very high EERs above 35\%. In particular, T10 and T22 had slightly higher EERs than 50\%, the chance level, and hence it is expected that T10 and T22 have slightly `emphasized' speaker characteristics compared to real target speakers. T23, T29, and T07 also had EERs between 45\% and 48\% and were higher than T27, T13, and T33.

\subsection{Analysis of prediction results for each language spoken by target speakers}

Task 2 of VCC 2020 is a cross-lingual VC task and the target speaker's speech contains utterances in German, Finnish, or Mandarin. As reported in \cite{vcc2020summary}, this factor affected the performance of the VC systems built by challenge participants. The converted audio files for German target speakers had the highest naturalness and speaker similarity, while those for Mandarin had the lowest. Therefore, we want to confirm whether the objective metrics can capture such score variations and predict the listening test scores for each language.   

A breakdown of the Pearson correlation coefficients per language is provided in Table \ref{tab:correlation-breakdown}. The top and bottom tables show correlations with the subjective scores from the English and Japanese listeners, respectively. The results for individual submissions are shown in Appendix \ref{appendix:language}.

As we can see, the correlations of the ASV metrics and ASR WER were similar and stable across the three languages. Again, MOS ratings were correlated with ASR WER and speaker similarity ratings were well correlated with ASV metrics. We can also see that the MOS ratings done by Japanese listeners had weaker correlations with ASR WER than those done by English listeners.

\begin{table*}[t!]
  \centering
  \caption{Coefficients ($p$-values) of multiple linear regression models that use ASR WER, MOSNet predictions, ASV EER, and countermeasure EER as inputs. Top: subjective scores from \textbf{English listeners}. Bottom: subjective scores from \textbf{Japanese listeners}.}
    \label{tab:regression}
    \footnotesize
    \centering
    \begin{tabular}{|c|c c c c c|c|c|c|}
        \hline
        \textbf{\makecell{Subjective\\score (ENG)}} & \textbf{Intercept} &  \textbf{\makecell{MOSNet\\(asvspoof19)}} & \textbf{\makecell{ASR\\WER (\%)}} & \textbf{\makecell{ASV \\EER (\%)}} & \textbf{\makecell{Countermeasure\\EER (\%)}} & \textbf{\makecell{Multiple\\R-Squared}} & \textbf{\makecell{Adjusted\\R-squared}} & \textbf{Significance F}\\
        \hline\hline
            Task 1 MOS & \makecell{1.713 \\ ($p=$0.054)} & \makecell{0.258 \\ ($p=$0.333)} & \makecell{$-$0.021 \\ \textbf{($p<$0.001)}} & \makecell{0.024 \\ \textbf{($p<$0.001)}} & \makecell{$-$0.002 \\ ($p=$0.654)} & 0.92 & 0.81 & $<$0.001 \\
            
            Task 1 SIM & \makecell{1.696 \\ ($p=$0.006)} & \makecell{0.062 \\ ($p=$0.722)} & \makecell{$-$0.003 \\ ($p=$0.146)} & \makecell{0.026 \\ \textbf{($p<$0.001)}} & \makecell{0.006 \\ ($p=$0.108)} & 0.92 & 0.83 & $<$0.001 \\
            
            Task 2 MOS & \makecell{0.619 \\ ($p=$0.357)} & \makecell{0.724 \\ \textbf{($p=$0.001)}} & \makecell{$-$0.021 \\ \textbf{($p<$0.001)}} & \makecell{0.014 \\ ($p=$0.049)} & \makecell{0.003 \\ ($p=$0.668)} & 0.88 & 0.74 & $<$0.001 \\
            
            Task 2 SIM & \makecell{1.782 \\ \textbf{($p<$0.001)}} & \makecell{0.038 \\ ($p=$0.617)} & \makecell{$-$0.002 \\ ($p=$0.254)} & \makecell{0.026 \\ \textbf{($p<$0.001)}} & \makecell{0.002 \\ ($p=$0.538)} & 0.91 & 0.80 & $<$0.001 \\
        \hline
    \end{tabular} \\
    \vspace{3mm}
    \centering
    \begin{tabular}{|c|c c c c c|c|c|c|}
        \hline
        \textbf{\makecell{Subjective\\score (JPN)}} & \textbf{Intercept} &  \textbf{\makecell{MOSNet\\(asvspoof19)}} & \textbf{\makecell{ASR\\WER (\%)}} & \textbf{\makecell{ASV \\EER (\%)}} & \textbf{\makecell{Countermeasure\\EER (\%)}} & \textbf{\makecell{Multiple\\R-Squared}} & \textbf{\makecell{Adjusted\\R-squared}} & \textbf{Significance F}\\
        \hline\hline
            Task 1 MOS & \makecell{1.024 \\ ($p=$0.239)} & \makecell{0.364 \\ ($p=$0.175)} & \makecell{$-$0.011 \\ \textbf{($p=$0.001)}} & \makecell{0.024 \\ \textbf{($p<$0.001)}} & \makecell{$-$0.002 \\ ($p=$0.765)} & 0.89 & 0.76 & $<$0.001 \\
            
            Task 1 SIM & \makecell{1.439 \\ ($p=$0.017)} & \makecell{0.132 \\ ($p=$0.453)} & \makecell{$-$0.001 \\ ($p=$0.475)} & \makecell{0.024 \\ \textbf{($p<$0.001)}} & \makecell{0.006 \\ ($p=$0.097)} & 0.91 & 0.81 & $<$0.001 \\
            
            Task 2 MOS & \makecell{-0.040 \\ ($p=$0.953)} & \makecell{0.833 \\ \textbf{($p<$0.001)}} & \makecell{$-$0.014 \\ \textbf{($p<$0.001)}} & \makecell{0.014 \\ ($p=$0.046)} & \makecell{0.006 \\ ($p=$0.395)} & 0.85 & 0.68 & $<$0.001 \\
            
            Task 2 SIM & \makecell{1.776 \\ \textbf{($p<$0.001)}} & \makecell{0.059 \\ ($p=$0.483)} & \makecell{$-$0.003 \\ ($p=$0.063)} & \makecell{0.023 \\ \textbf{($p<$0.001)}} & \makecell{0.004 \\ ($p=$0.196)} & 0.88 & 0.74 & $<$0.001 \\
        \hline
    \end{tabular}
\end{table*}

\subsection{Can we predict subjective evaluation results better by using combinations of the objective metrics?}

Next, we carried out multiple linear regression analysis to determine whether the prediction accuracy on subjective scores could be improved by combining several objective metrics. Since the ASV metrics reported in previous subsections have ``multicollinearity'', we selected ASV EER only for this analysis. Also, we chose MOSNet (asvspoof19) over MOSnet (vcc18) since the former had higher correlation with the subjective scores (as shown in Table~\ref{tab:correlation}). The estimated coefficients and the statistics are listed in Table~\ref{tab:regression}, where the top and bottom parts show subjective scores from English and Japanese listeners, respectively.

From the table, by comparing the adjusted R-squared values and the Pearson correlation coefficients of Table \ref{tab:correlation}, we can see that the prediction accuracy on subjective quality rating scores can be improved by combining multiple objective metrics for all of Task 1 and for the quality rating of Task 2. The significant explainable variables for MOS were ASV EER and ASR WER for Task 1 and MOSNet (asvspoof19) and ASR WER for Task 2. This was the case for both the English and Japanese listeners' scores. However, this was not the case for the subjective speaker similarity rating, and only the coefficient for ASV EER was statistically significant. This is presumably because the ASV EER itself had sufficiently high correlations. This finding is consistent with the correlation analysis discussed in the previous section. 

We can also see that the MOS score regression models in Task 2 had lower adjusted R-squared values (0.74 and 0.68) compared to the other regression results. This clearly suggests that predicting the naturalness score in the cross-lingual setting is harder, and needs to be explained by more factors.

\subsection{Discussion}

The results of our analysis demonstrated that MOS ratings were correlated with ASR WER or MOSNet (asvspoof19) and speaker similarity ratings were well correlated with ASV metrics, and that they were complementary for quality rating. 

Contrary to our expectations, only the ASV and ASR models trained using human speech data became important explainable variables for predicting subjective ratings, and the CM models using synthesized speech or converted speech did not. This might be due to the many new types of waveform generation methods adopted by the challenge participants. As reported in \cite{vcc2020summary}, ten types of vocoders were used for VCC 2020, many of which were not included in the training databases used for the CM model. 

We should point out that the correlations reported in this paper are at the system level and as such represent neither the quality nor the speaker similarity of individual sentences. Investigating sentence-level score predictions will be the focus of future work.

\section{Spoofing Performance Assessment}

In this section, we address our second question, ``Which VC technology has the highest spoofing risk for ASV and CM?'' If we can answer this question and identify which VC methods the current ASV and CM systems are most vulnerable to, the speaker recognition community can use our results as a guideline to efficiently prepare additional training data to enhance the robustness of ASV and CM systems. Once spoofed data produced by the missing VC technologies is added to training databases for the CM systems and the VC technologies become ``known'' from the CM perspective, their discrimination normally becomes much easier. Therefore, we focus on the ASV and CM metrics in this section. 

\subsection{t-DCF-based spoofing performance analysis for each submitted VC system}

The ASV and CM results in Tables \ref{tab:obj_Task1} and \ref{tab:obj_Task2} were reported in isolation from each other. For ``spoofing performance'' assessment, we consider the ASV and CM results \emph{jointly}. Specifically, we envision a cascaded (tandem) system \cite{Kinnunen2020-tandem} where a CM system is placed before the ASV, with the aim of preventing spoofing attacks from reaching the ASV system. VC audio files may be rejected by a CM system if the audio contains detectable artifacts. Even if VC audio files are passed on to the CM system, they may still be rejected by the ASV if their speaker similarity is not close enough to the target speakers. Further, while our primary interest is the performance degradation due to falsely accepted VC spoofing attacks, the tandem system is also prone to two other types of errors. First, it may accept another similar-appearing human user (a \emph{non-target}) as a target. Second, either CM or ASV may reject the actual target user. Using an overly sloppy CM or ASV system leads to compromised security, while conversely, an overly aggressive CM (or ASV) leads to reduced user convenience. To assess the joint performance of CM and ASV, we adopt a new metric called (minimum) \emph{tandem detection cost function} (t-DCF) \cite{Kinnunen2020-tandem}. It combines all the ASV and CM system errors into a single cost value for each submitted VC system.  

Unlike the parameter-free metrics considered above, t-DCF is a parameterized cost that makes the modeling assumptions of an envisioned operating environment (application) explicit. A desired security-convenience trade-off is specified through \emph{detection costs} assigned to erroneous system decisions and \emph{prior probabilities} assigned to the commonality of targets, non-targets, and spoofing attacks. The ASV threshold is set to the EER point on bona fide samples. Following the notations and parameter constraints in \cite{Kinnunen2020-tandem}, we assign costs $C_\text{miss}=1$, $C_\text{fa}=C_\text{fa,spoof}=10$, and priors $\pi_\text{spoof}=0.50$, $\pi_\text{tar}=(1 - \pi_\text{spoof})\times 0.99 \approx 0.4950$, $\pi_\text{non}=(1 - \pi_\text{spoof})\times 0.01 \approx 0.0050$. This is representative of a high-security user authentication application (e.g., access control) where target users and spoofing attacks are almost equally likely to occur, while nontarget users are rare. False acceptances (whether of nontargets or VC attacks) incur a ten-fold cost relative to false rejections. The higher the t-DCF value, the more detrimental the VC attack. The maximum value of 1.0 indicates an attack that renders the tandem system useless.

\begin{table}
  \centering
  \caption{Minimum t-DCF for each system of VCC 2020. Red cells indicate top-5 systems for each task.}
    \label{tab:tdcf}
     \resizebox{8.0cm}{!}{
    \begin{tabular}{|c|c|c||c|c|c|}
        \hline
        \textbf{System}  & \textbf{Task 1} & \textbf{Task 2} & \textbf{System}  & \textbf{Task 1} & \textbf{Task 2} \\
        \hline\hline
T01 &  0.73542  &  -- &  T18 & 0.70372 & 0.81145\\
T02 & 0.85274   & 0.70888 &  T19 & 0.8743 & \cellcolor{red!25} 0.90471\\
T03 & 0.01467   & 0.01467 &  T20 & 0.85301 & 0.77249\\
T04 & 0.88342   &  --  &  T21 & 0.86755 & --\\
T05 &  -- & 0.60904 &  T22 & 0.86204 & \cellcolor{red!25} 0.93512\\
T06 & \cellcolor{red!25} 1.0000  & 0.72722 &  T23 & 0.8297 & \cellcolor{red!25} 0.9037\\
T07 & 0.87227   & 0.9033 &  T24 & 0.76482 & 0.79092\\
T08 & \cellcolor{red!25} 1.00000   & \cellcolor{red!25} 1.00000 &  T25 & 0.85402 & 0.85048\\
T09 & 0.25987   & 0.29213 &  T26 & 0.71041 & 0.53263\\
T10 & 0.87126   & \cellcolor{red!25} 0.91282 &  T27 & 0.80151 & 0.84287\\
T11 & 0.87531   & 0.88646 &  T28 & \cellcolor{red!25} 0.91214 & 0.82598\\
T12 & \cellcolor{red!25} 1.00000  & 0.84693 &  T29 & 0.83375 & 0.87311\\
T13 & 0.88646  & 0.79685 &  T30 & 0.04508  & 0.09695\\
T14 & \cellcolor{red!25} 0.91708   &  --     &  T31 & 0.84069 & 0.70379\\
T15 & -- & 0.8805  &  T32 & 0.80942 & 0.76208\\
T16 & 0.87633   & 0.88818 &  T33 & 0.78095 & 0.83375\\
T17 & 0.87734  & --      &  --      & --     &--\\
        \hline
    \end{tabular}
  }
\end{table}

Table~\ref{tab:tdcf} shows the minimum t-DCF performance for each team of VCC 2020, where the top-5 systems showing the highest minimum t-DCF values are highlighted for both tasks. Among these top-5 systems, team T08 reached the maximum possible cost (1.00) in both tasks. Table \ref{tab:obj_Task1} reveals that T08 yielded nearly zero ASV EER (0.50\%)---i.e., it did not succeed in fooling ASV. At the same time, however, the corresponding CM EER was very high (37.97\%), indicating difficulties in detecting this spoofing attack. This might be due to a lacking spoofing artifacts in T08, issues with CM generalization, or both. 

The t-DCF was high in this case due to the non-accurate CM that falsely rejected many target utterances. Since the ASV would have rejected this unsuccessful attack with high probability, \emph{it might have been better to not use any CM system at all}\footnote{The mathematical properties of t-DCF \cite{Kinnunen2020-tandem} imply that if the spoof false acceptance rate (SFAR) of the ASV system is 0, the optimal countermeasure is \emph{no countermeasure}. Interested readers are directed to \cite[Eq. (10)]{Kinnunen2020-tandem}, where $C_2=0$ whenever $P_\text{fa,spoof}^\text{asv}=0$. A CM that minimizes Eq. (10) in this case must use a detection threshold $\tau^\text{cm}=-\infty$, aka `accept everything' or `no countermeasure'.}---certainly not a low-performing one. This general pattern (low ASV EER \%, relatively high CM EER \%) also held for T06, T12, and T14 in Task 1. For T28, however, the high t-DCF value was due to the relatively high EER for both CM and ASV. 

In Task 2, apart from T08, we find a similar explanation for high t-DCF values in the top-5 VC systems. Taking T10 as an example, the ASV EER was 45.55\% and CM EER was 49.81\%; in other words, T10 fooled the ASV nearly perfectly. At the same time, the CM could do no better than random guessing in detecting this attack. Thus, T10 is a highly effective attack that is difficult to detect (by our CM). Careful examination of Table \ref{tab:obj_Task1} reveals this pattern (high EERs on both ASV and CM) for T19, T22, and T23 as well.

\begin{table}
\caption{\label{table_tDCF_sys} Details of top-performing VC systems in terms of minimum t-DCF as a spoofing threat.}
\centering
\footnotesize
\begin{tabular}{|c|l|l|} 
\hline
\multicolumn{3}{|c|}{\textbf{Task 1}}\\ \hline                     {\textbf{Team ID }} & {\textbf{VC model~}} & {~\textbf{Vocoder}}\\ 
\hline
T06 & StarGAN & WORLD\\
T08 & VTLN + Spectral differential & WORLD\\
T12 & ADAGAN & AHOcoder\\
T14 & One-shot VC & NSF\\
T28 & Tacotron & WaveRNN\\
\hline
\hline
\multicolumn{3}{|c|}{\textbf{Task 2}}\\ \hline                     {\textbf{Team ID }} & {\textbf{VC model~}} & {~\textbf{Vocoder}}\\ 
\hline
T08 & VTLN + Spectral differential & WORLD\\
T22 & ASR-TTS (Transformer) & Parallel WaveGAN\\
T10 & PPG-VC (LSTM) & WaveNet\\
T19 & VQVAE & Parallel WaveGAN\\
T23 & CycleVAE & WaveNet\\
\hline
\end{tabular}
\end{table}

Next, we take a closer look at the top-5 systems to determine which VC approaches show a potential threat for spoofing in terms of t-DCF. Table~\ref{table_tDCF_sys} shows the details of the VC approaches used in the top-5 systems for spoofing assessment under the t-DCF measure, whose details are taken from our VCC 2020 summary paper~\cite{vcc2020summary}. With this analysis we try to highlight some of the VC models and vocoders that have a potentially high spoofing threat. As expected, unseen VC methods that are not included in the training database (e.g., GAN variants (StarGAN, AdaGAN, Parallel WaveGAN) and VTLN) had the highest t-DCF values. These VC methods should be prioritized as attack methods added to countermeasure databases in the future.

\subsection{Discussion}

Do the VC methods featured in the VCC 2020 impose a spoofing risk? Yes, they certainly do. One useful reference is the ASV performance on natural human samples---more specifically, EER on target-nontarget discrimination. These EERs were 0.50\% and 0.80\% for Tasks 1 and 2, respectively. Most VC systems increased these numbers, most of them substantially. This is not news to the ASV community as such. Our spoofing performance assessment through t-DCF clearly highlights the importance of improving both ASV and CM technology. The VCC 2020 not only featured a battery of new VC techniques but also facilitated an initial CM performance benchmarking on a new type of multilingual data. Further research is needed to analyze the combined effect of VC methods and language. Substantial future work remains in improving the generalizability of CM techniques across diverse data conditions.

\section{Conclusion}

This work summarizes the predictions of subjective ratings and spoofing assessments with objective assessments performed at the latest VCC 2020. We considered five different objective assessments based on ASV, neural speaker embedding, spoofing countermeasure, predicted MOS, and ASR. The correlations of objective assessments computed with subjective ratings indicate that the ASV, neural speaker embedding, and ASR had high correlations, which suggests the possibility of predicting the subjective ratings. Further, we found that the ASV and ASR results were more effective than the predicted MOS and spoofing countermeasure for predicting the subjective test results using multiple linear regression models. This indicates the potential shift toward relying on the objective assessments over tedious listening tests for large-scale evaluations in the future. We performed further spoofing assessment on the submissions and identified the VC methods with a potentially high threat. However, this topic deserves future exploration as the performance highly depends on the coverage of the various VC methods included in the training data.

\section{Acknowledgements}
The authors thank Ms. Jennifer Williams of the University of Edinburgh for kindly providing a MOSNet model fine-tuned using the ASVspoof 2019 data. This work was partially supported by JST CREST Grants (JPMJCR18A6, VoicePersonae project, and JPMJCR19A3, CoAugmentation project), Japan, MEXT KAKENHI Grants (16H06302, 17H04687, 17H06101, 18H04120, 18H04112, 18KT0051, 19K24373), Japan, the National Natural Science Foundation of China (Grant No. 61871358), Programmatic Grant No. A1687b0033 from the Singapore Government’s Research, Innovation and Enterprise 2020 plan (Advanced Manufacturing and Engineering domain), and the Academy of Finland (project no. 309629).

\balance
\bibliographystyle{IEEEtran}
\bibliography{main}

\onecolumn

\newpage
\appendix

\section{Results of Individual Metrics} 

Here, we show the rankings and comparison of every team based on the various objective measures for both tasks discussed in Section~\ref{objEvalVCC2020}.  

\begin{figure}[h!]
    \centering
    \includegraphics[width=0.65\textwidth]{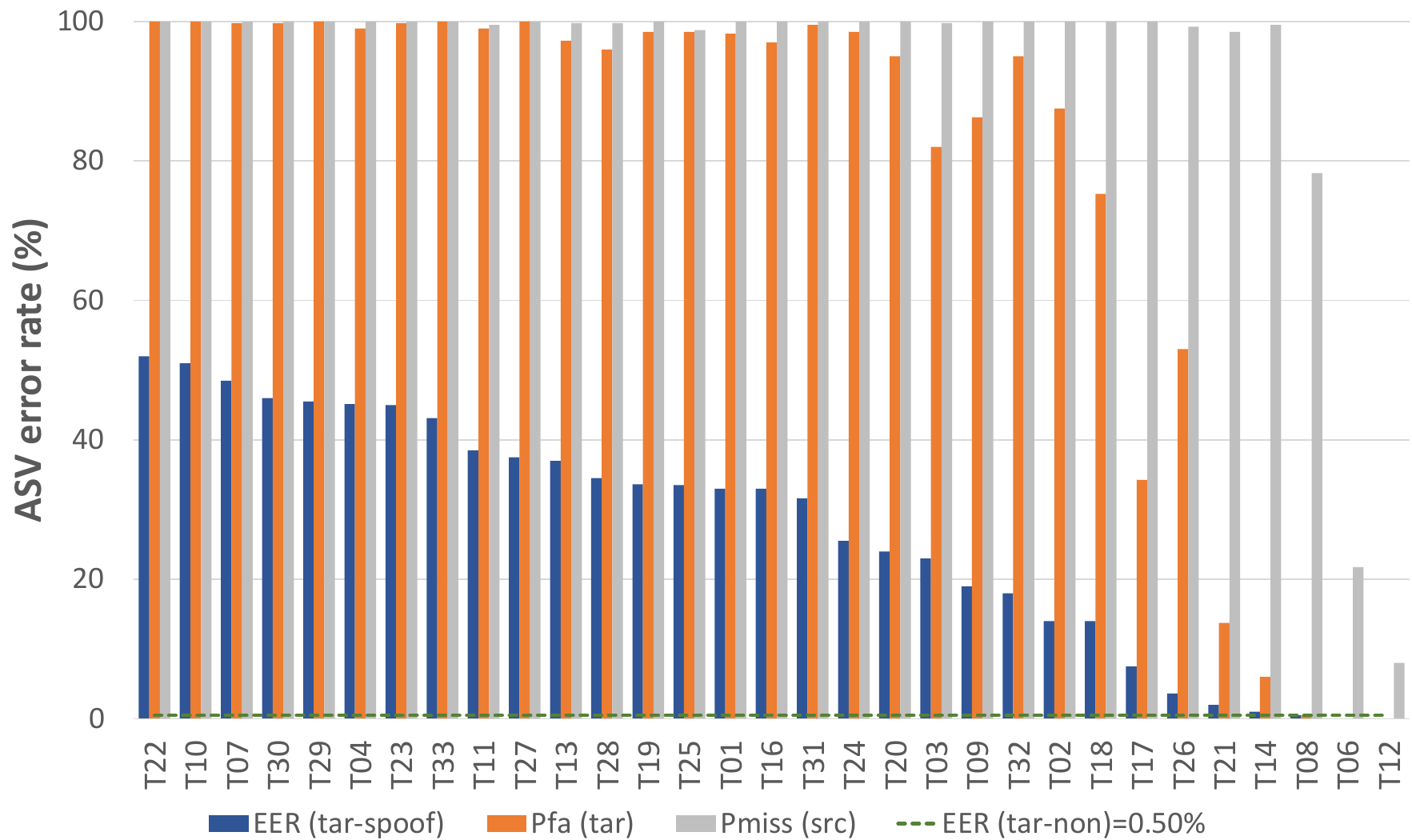} 
    \label{fig:subim1}
    \caption{Summary of ASV-based speaker similarity assessment for Task 1.}
\end{figure}

\begin{figure}[h!]
    \centering
    \includegraphics[width=0.65\textwidth]{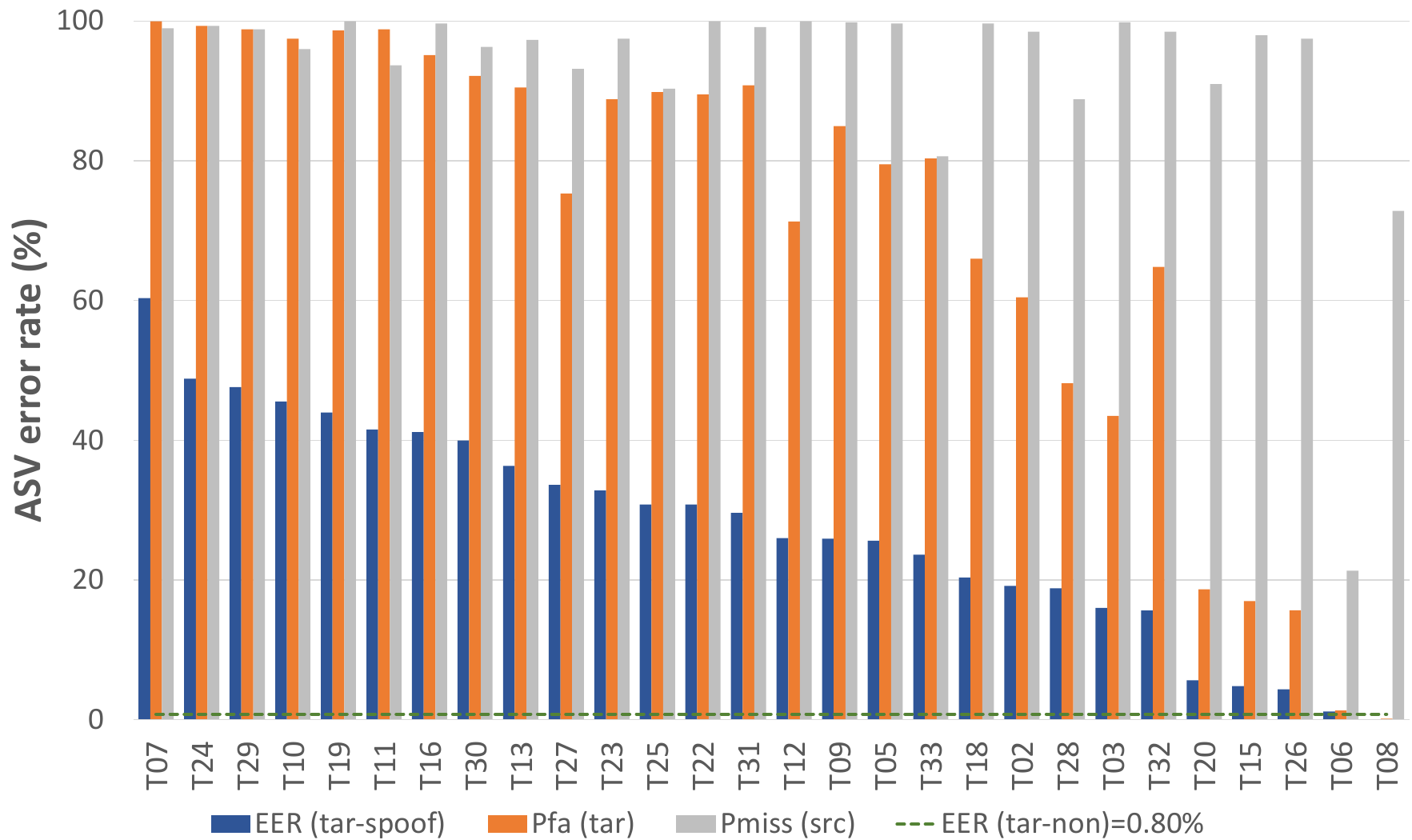}
    \label{fig:subim2}
    \caption{Summary of ASV-based speaker similarity assessment for Task 2.}
\end{figure}

\begin{figure}[h!]
	\centering
	 \includegraphics[width=0.7\textwidth]{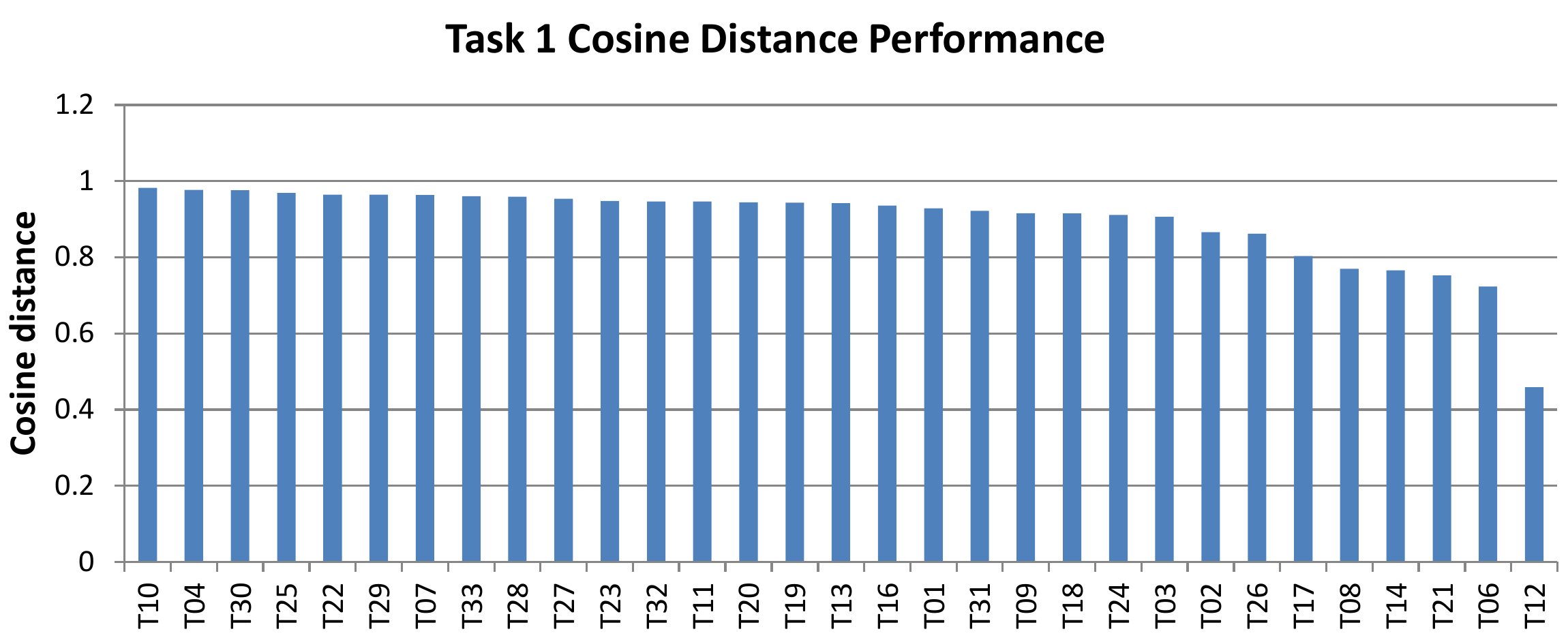}
	\caption{Summary of cosine distance-based neural speaker embedding for Task 1.}
	\label{fig:CosDist-task1}
\end{figure}

\begin{figure}[h!]
	\centering
	\includegraphics[width=0.68\textwidth]{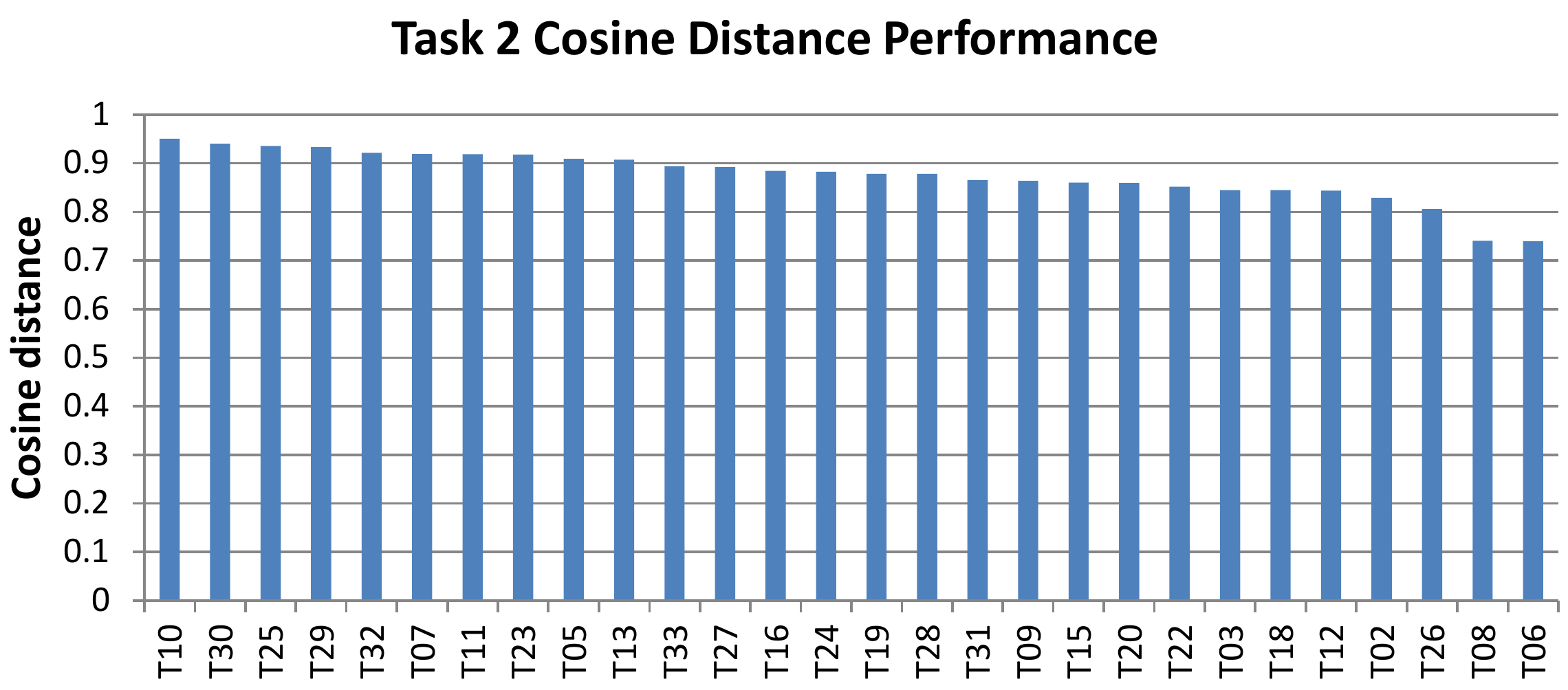}
	\caption{Summary of cosine distance-based neural speaker embedding for Task 2.}
	\label{fig:CosDist-task2}
\end{figure}

\begin{figure}[h!]
	\centering
	 \includegraphics[width=0.68\textwidth]{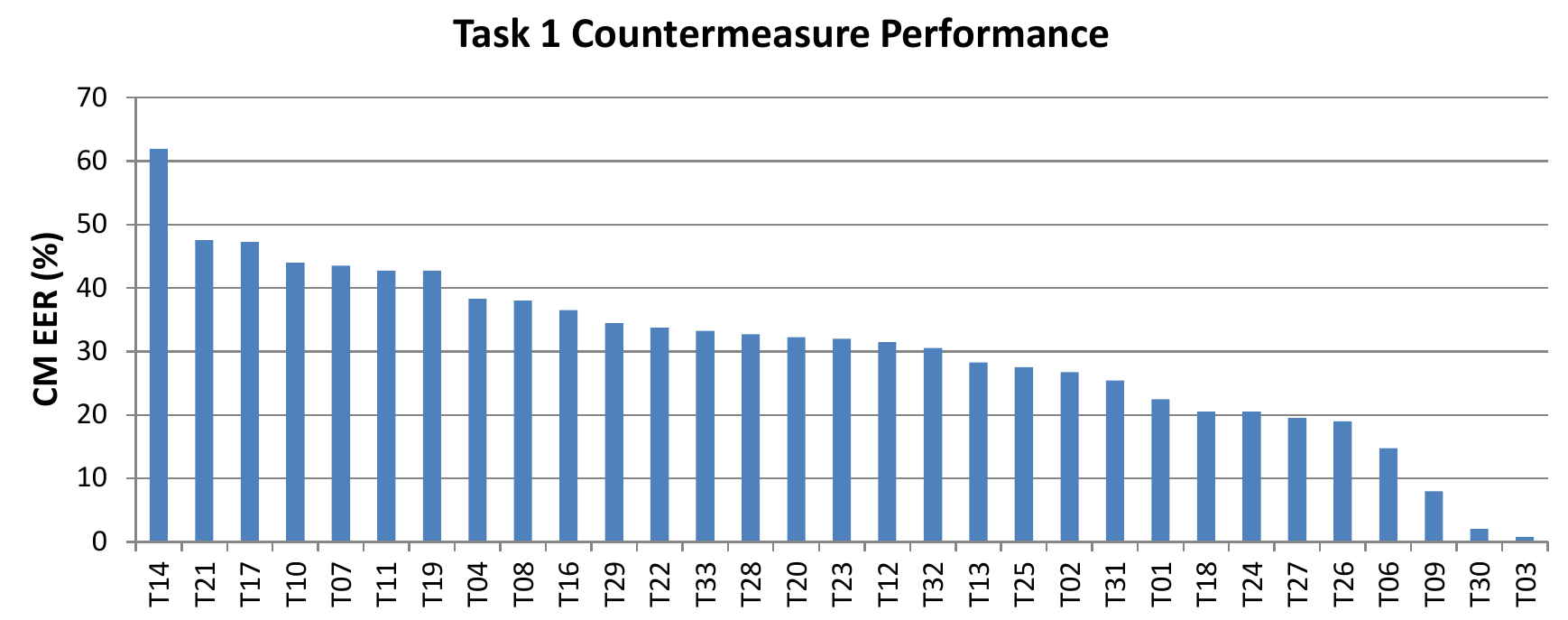}
	\caption{Summary of spoofing countermeasure EER (\%) for Task 1.}
	\label{fig:cm-bar-task1}
\end{figure}

\begin{figure}[h!]
	\centering
	\includegraphics[width=0.68\textwidth]{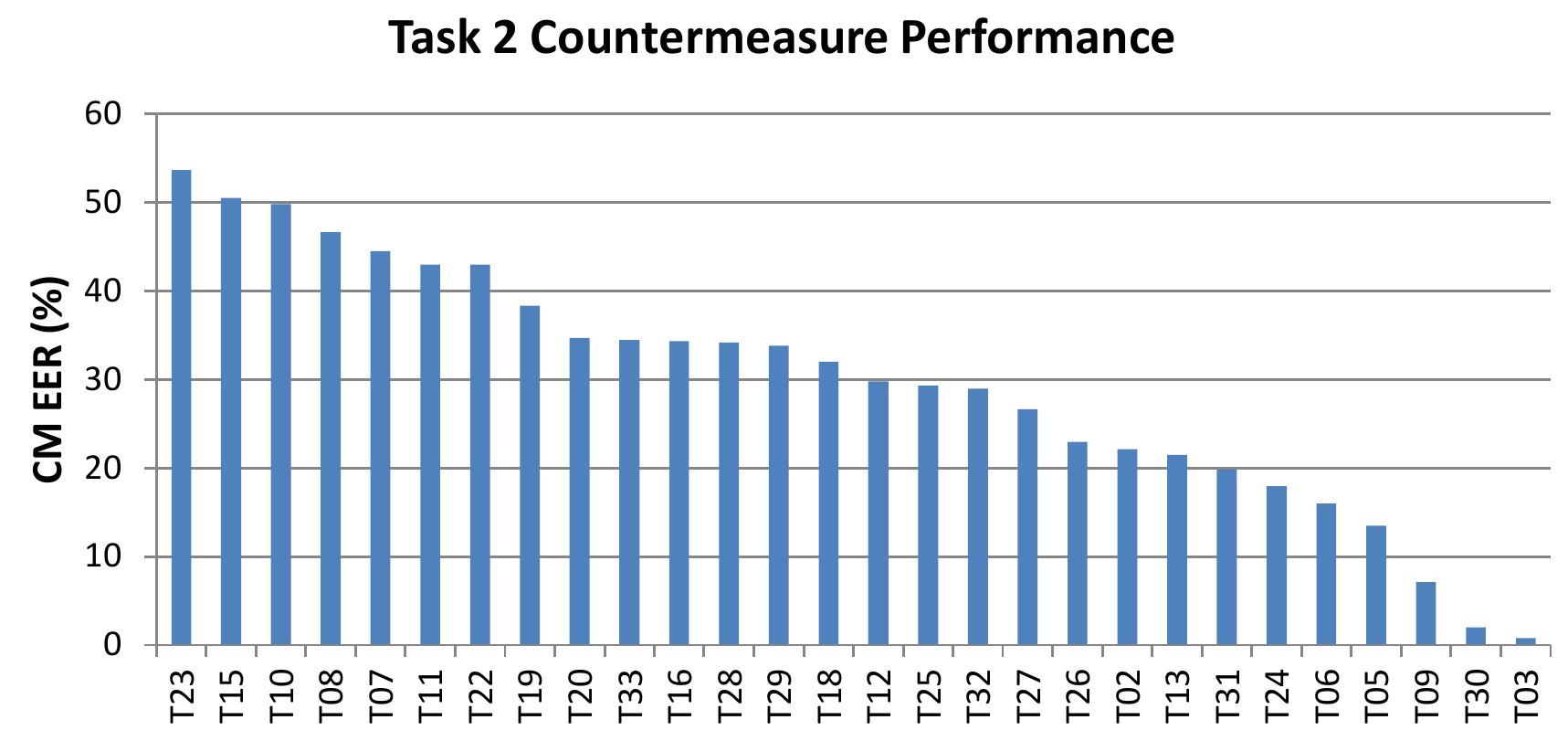}
	\caption{Summary of spoofing countermeasure EER (\%) for Task 2.}
	\label{fig:cm-bar-task2}
\end{figure}

\begin{figure}[h!]
	\centering
	 \includegraphics[width=0.68\textwidth]{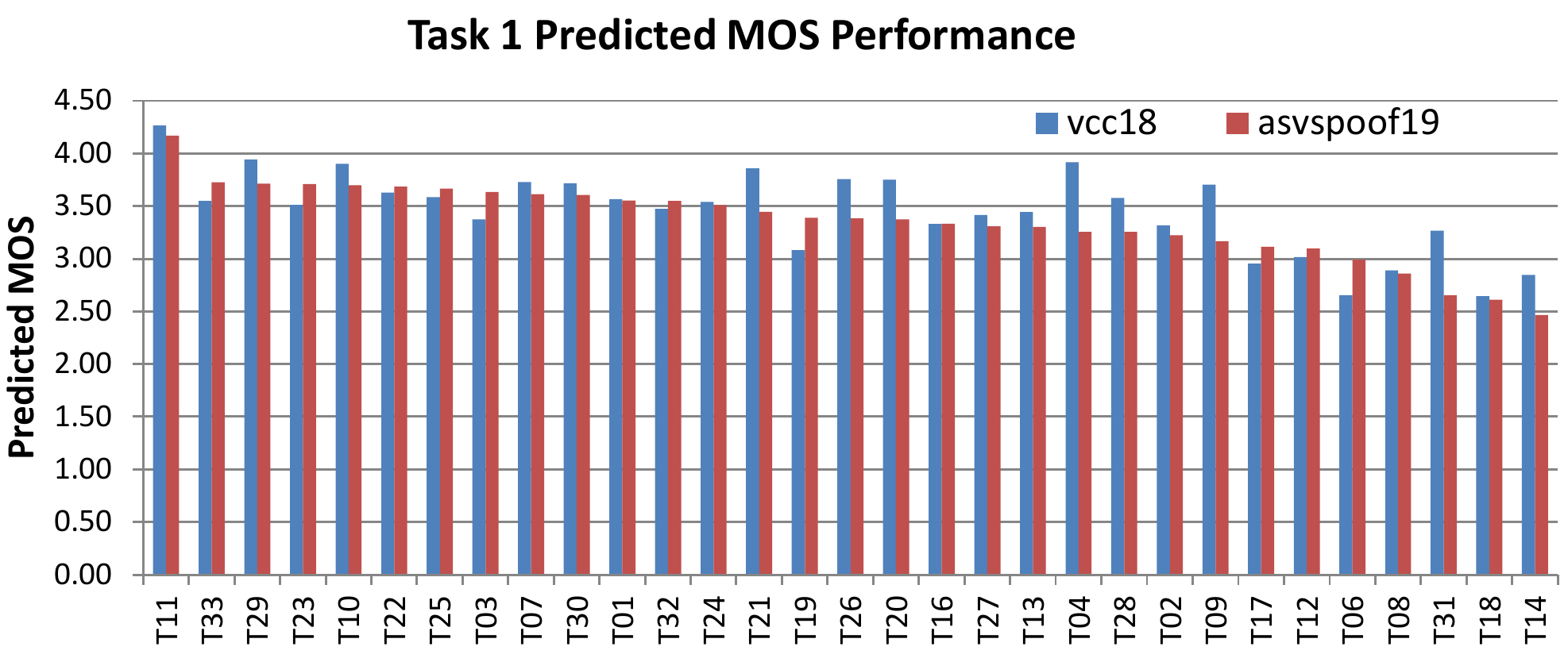}
	\caption{MOSNet predictions of all systems in Task 1.}
	\label{fig:mosnet-bar-task1}
\end{figure}
\begin{figure}[h!]
	\centering
	\includegraphics[width=0.68\textwidth]{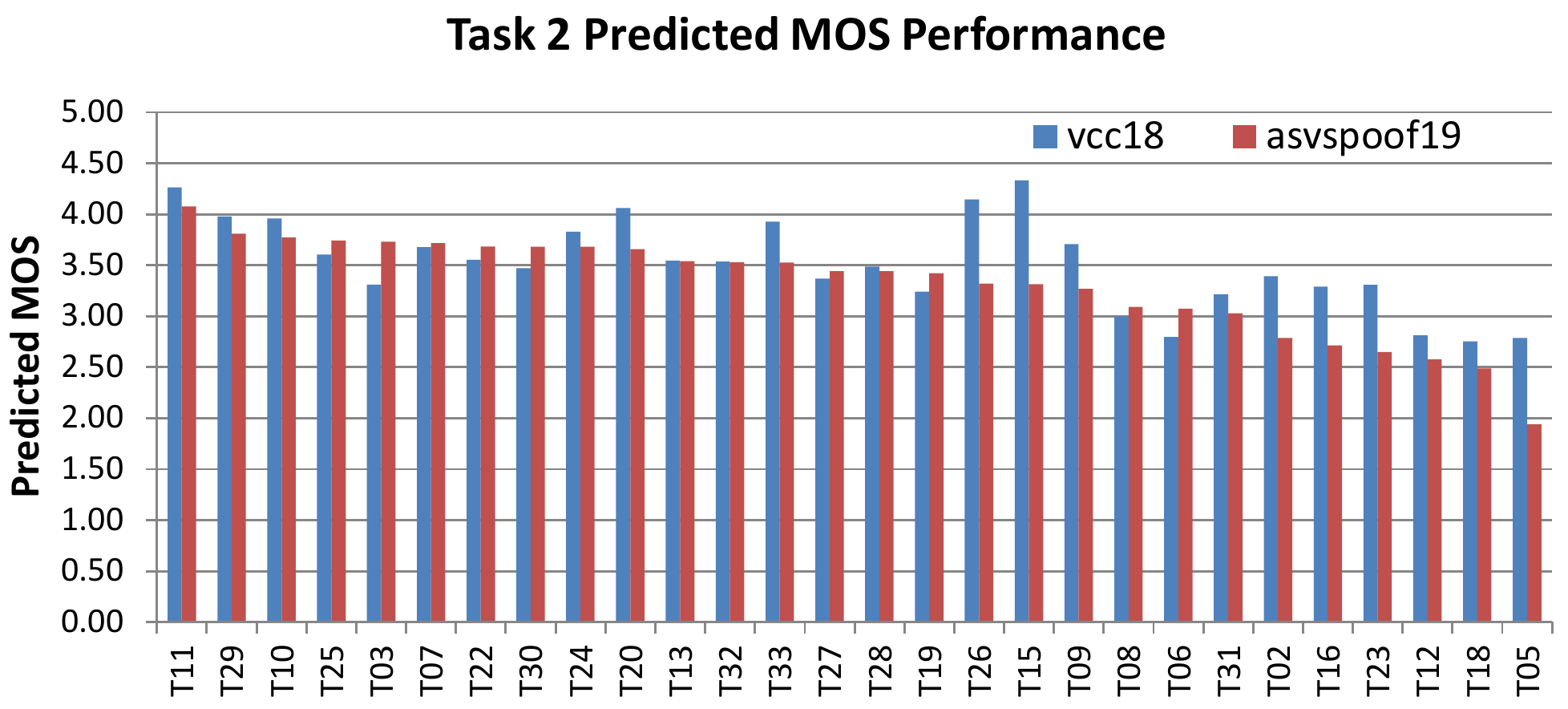}
	\caption{MOSNet predictions of all systems in Task 2.}
	\label{fig:mosnet-bar-task2}
\end{figure}

\begin{figure}[h!]
	\centering
	 \includegraphics[width=0.68\textwidth]{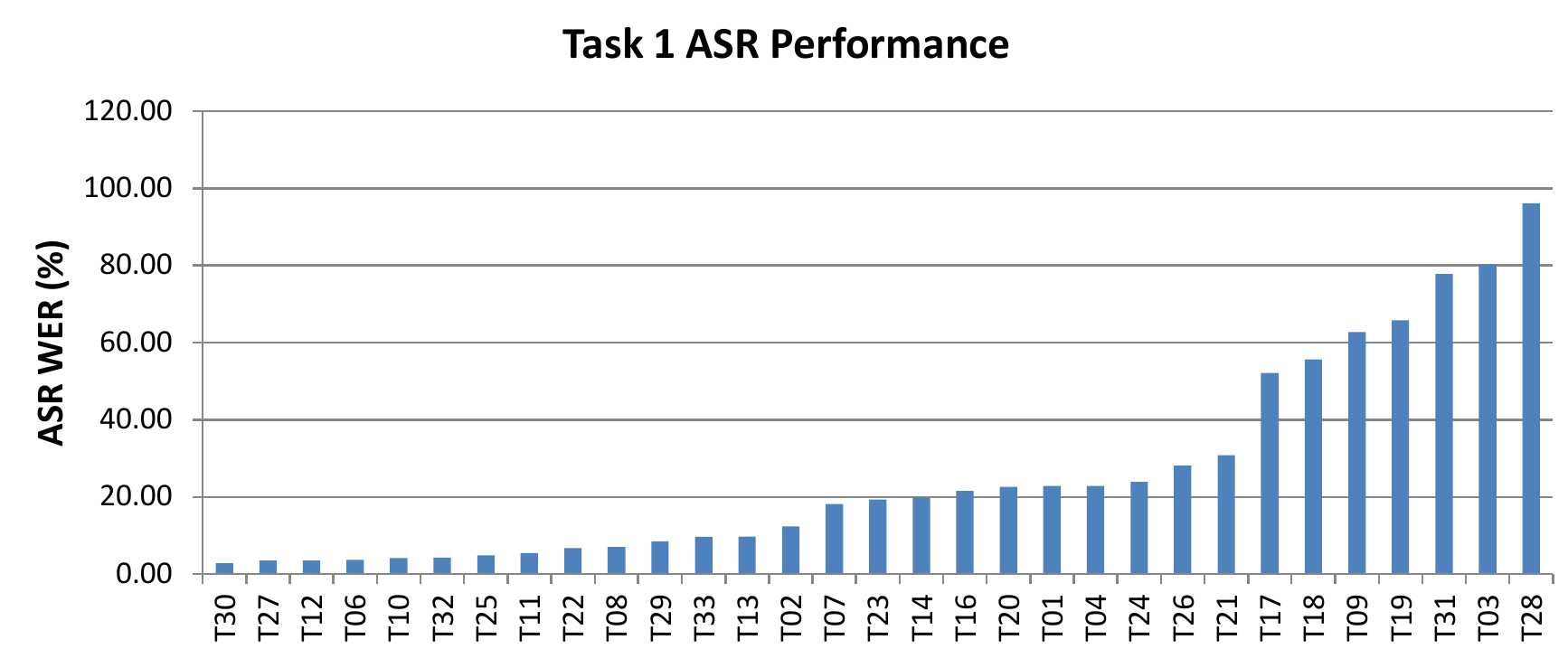}
	\caption{Summary of ASR WER (\%) for Task 1.}
	\label{fig:asr-bar-task1}
\end{figure}

\begin{figure}[h!]
	\centering
	\includegraphics[width=0.68\textwidth]{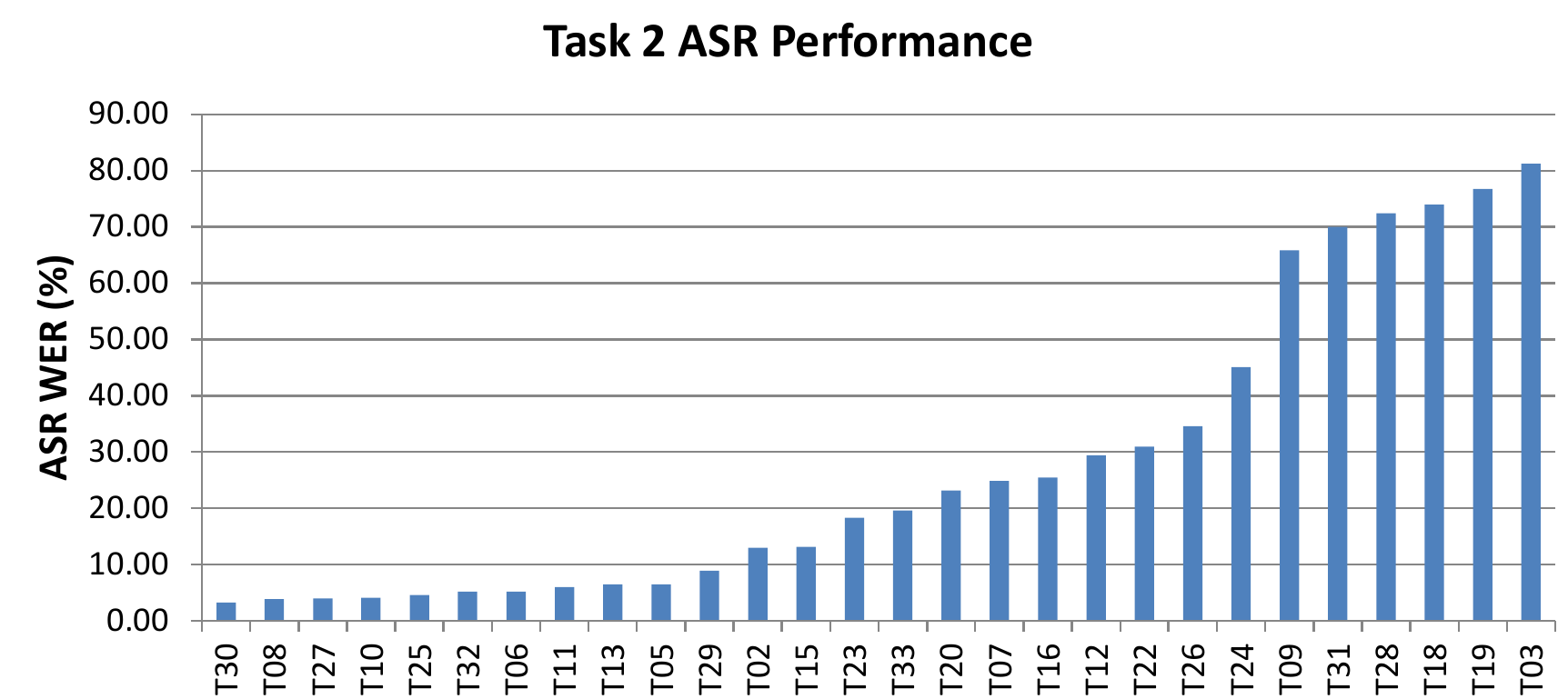}
	\caption{Summary of ASR WER (\%) for Task 2.}
	\label{fig:asr-bar-task2}
	\vspace{-4mm}
\end{figure}

\clearpage
\section{Objective Evaluation Results and Scatter Plots} 
\label{sub-vs-obj-scatter}

\begin{figure}[h!]
\begin{center}
\includegraphics[width=0.85\textwidth]{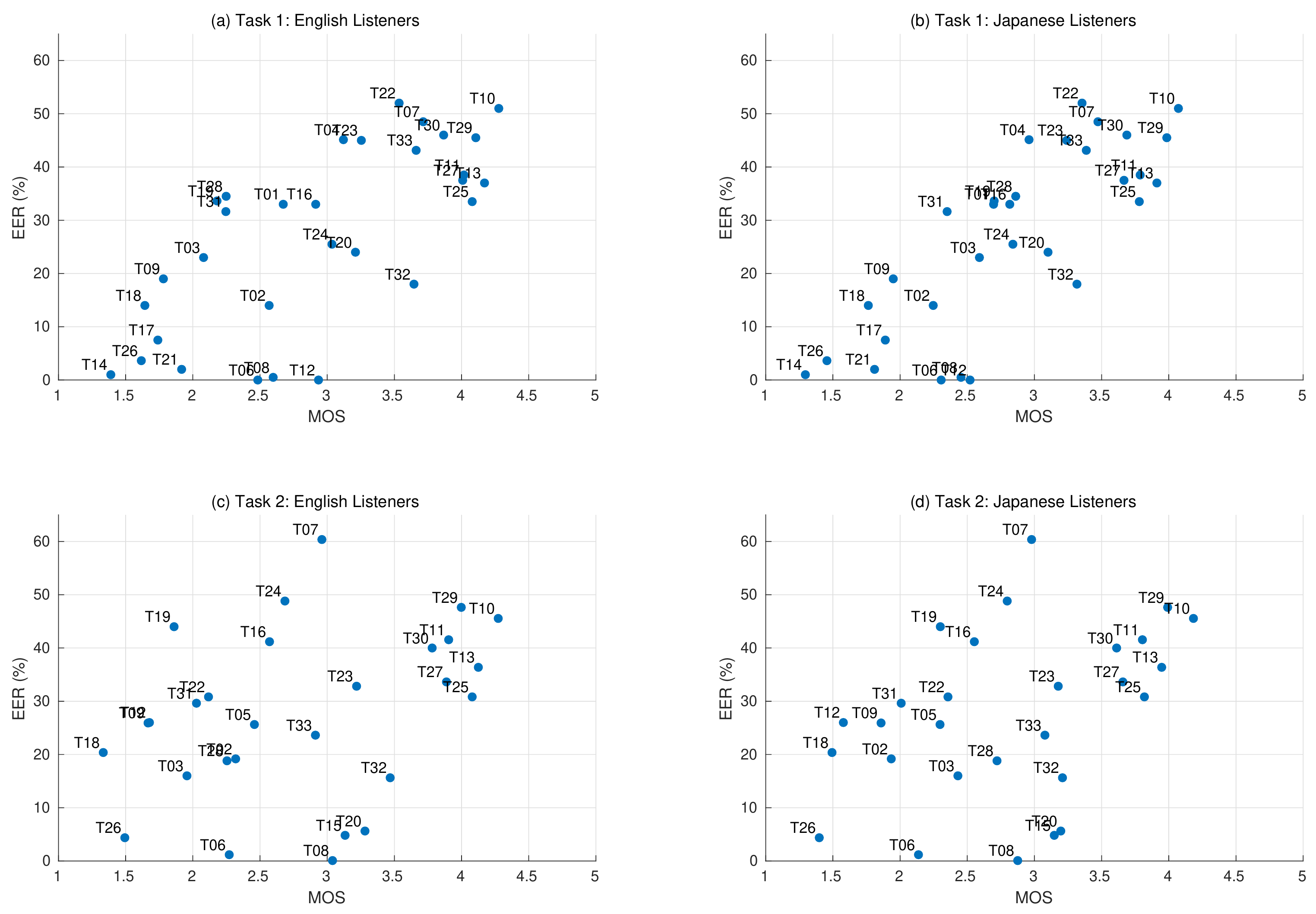}
\end{center}
\vspace{-5mm}
\caption{Scatter plots for ASV EER (\%) with subjective MOS.} 
\label{corr_ASV_EER_MOS}
\end{figure}

\begin{figure}[h!]
\begin{center}
\vspace{-8mm}
\includegraphics[width=0.85\textwidth]{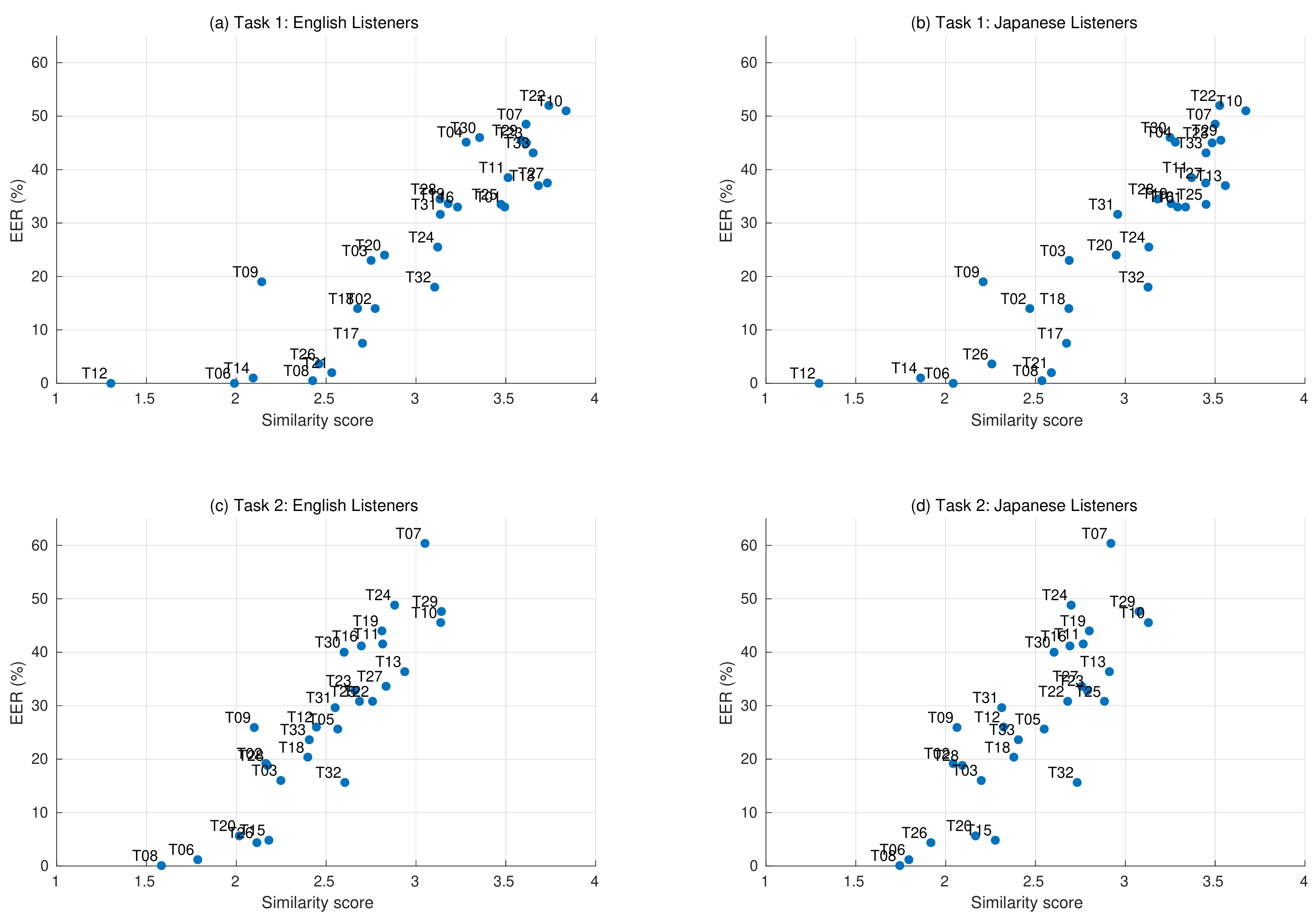}
\end{center}
\vspace{-5mm}
\caption{Scatter plots for ASV EER (\%) with subjective speaker similarity.} 
\label{corr_ASV_EER_SIM}
\end{figure}

\begin{figure}[h!]
\begin{center}
\includegraphics[width=0.85\textwidth]{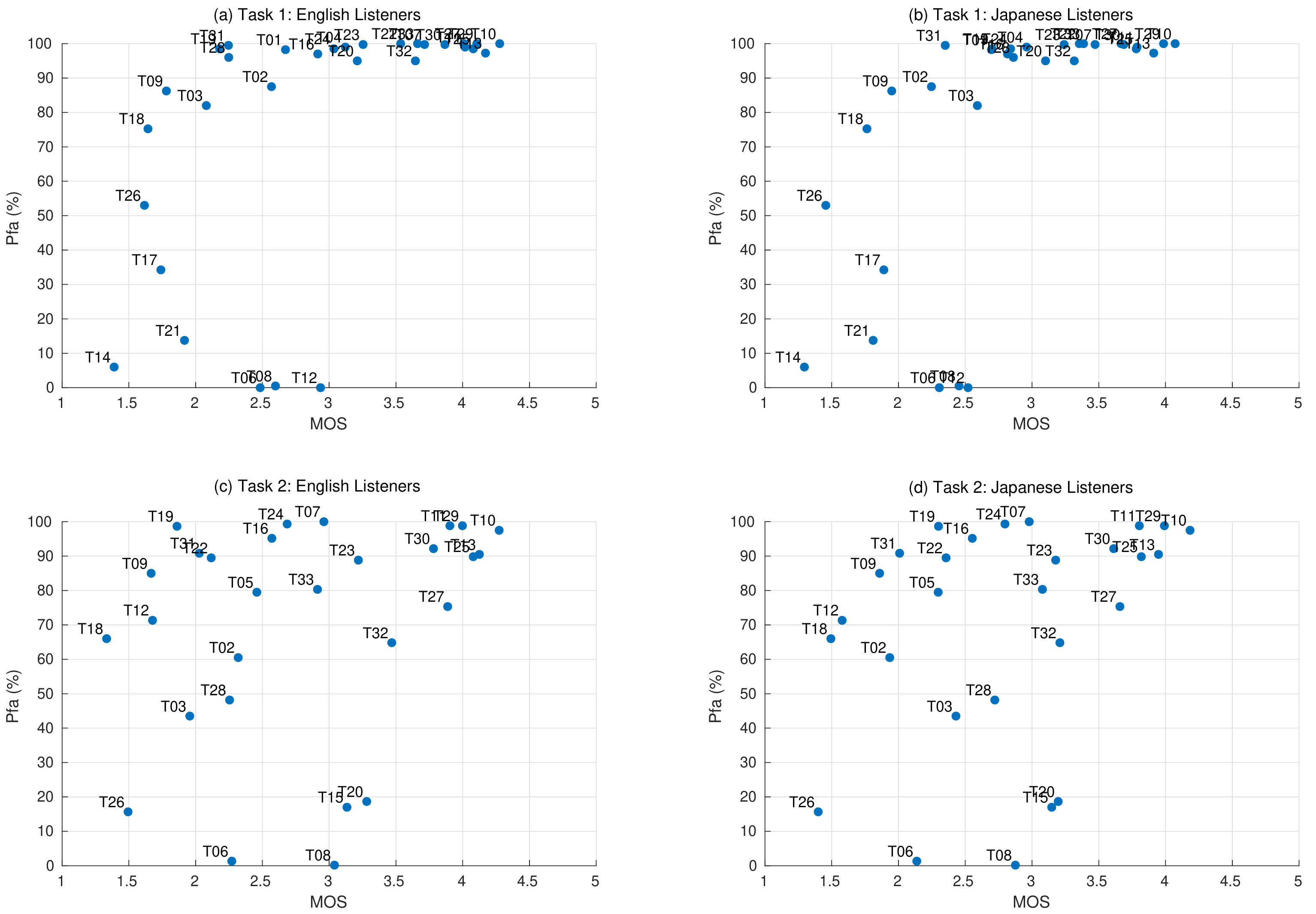}
\end{center}
\caption{Scatter plots for ASV Pfa (\%) with subjective MOS.} 
\label{corr_ASV_Pfa_MOS}
\end{figure}

\begin{figure}[h!]
\begin{center}
\includegraphics[width=0.85\textwidth]{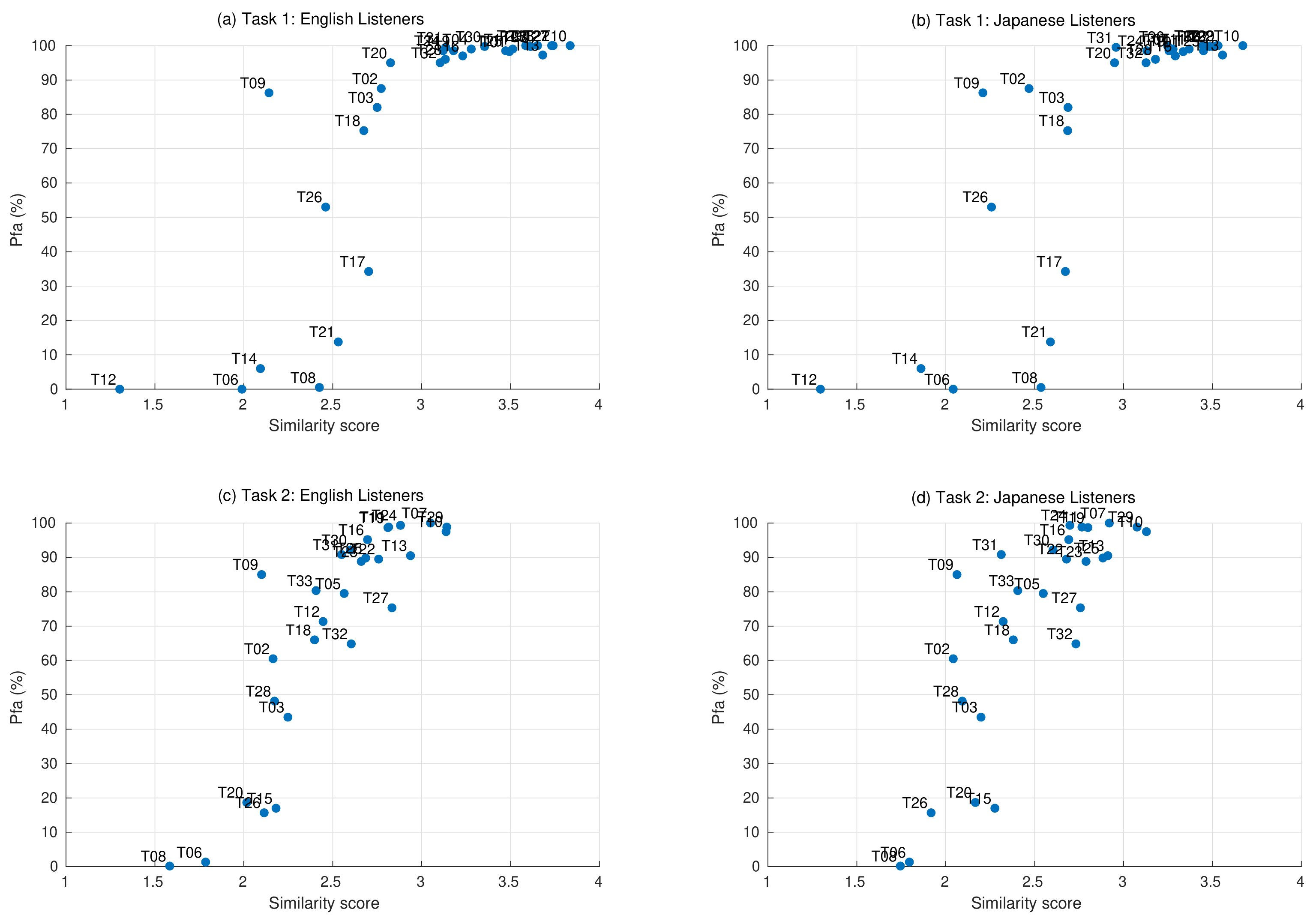}
\end{center}
\caption{Scatter plots for ASV Pfa (\%) with subjective speaker similarity.} 
\label{corr_ASV_Pfa_SIM}
\end{figure}

\begin{figure}[h!]
\begin{center}
\includegraphics[width=0.85\textwidth]{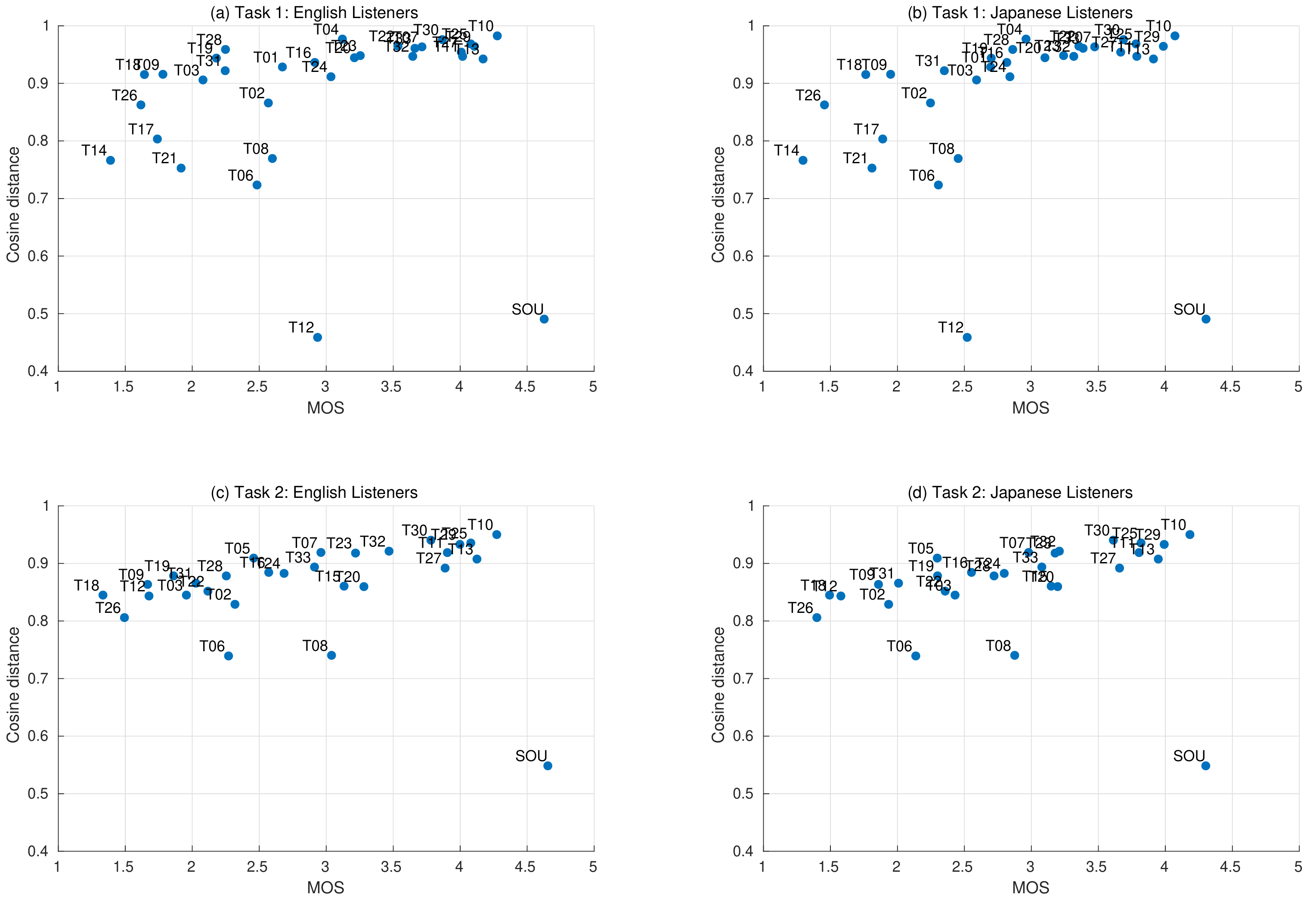}
\end{center}
\caption{Scatter plots for speaker embedding cosine distance with subjective MOS.} 
\label{corr_CDS_MOS}
\end{figure}

\begin{figure}[h!]
\begin{center}
\includegraphics[width=0.85\textwidth]{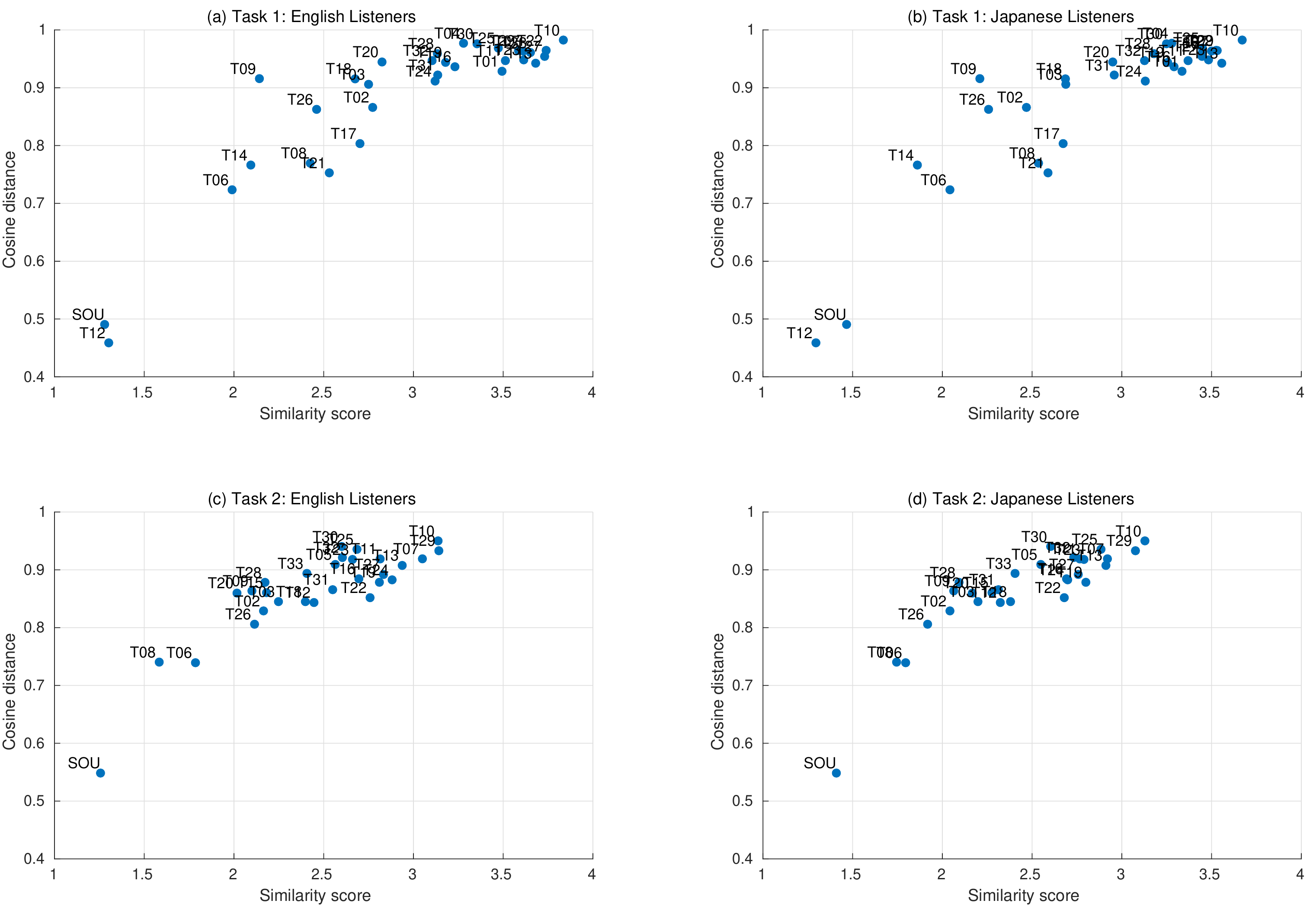}
\end{center}
\caption{Scatter plots for speaker embedding cosine distance with subjective speaker similarity.} 
\label{corr_CDS_SIM}
\end{figure}

\begin{figure}[h!]
\begin{center}
\includegraphics[width=0.85\textwidth]{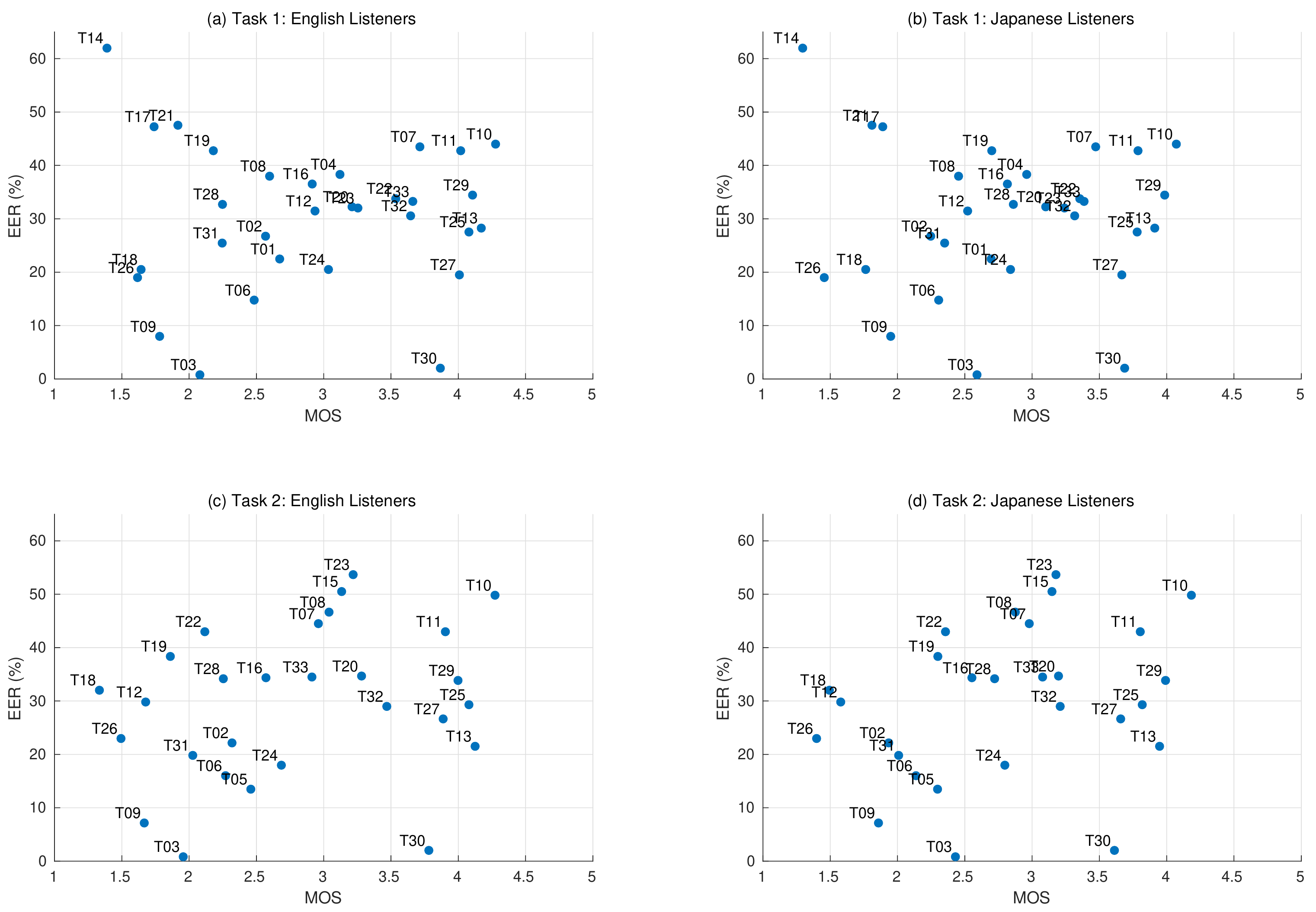}
\end{center}
\caption{Scatter plots for spoofing countermeasure EER (\%) with subjective MOS.} 
\label{corr_CM_EER_MOS}
\end{figure}

\begin{figure}[h!]
\begin{center}
\includegraphics[width=0.85\textwidth]{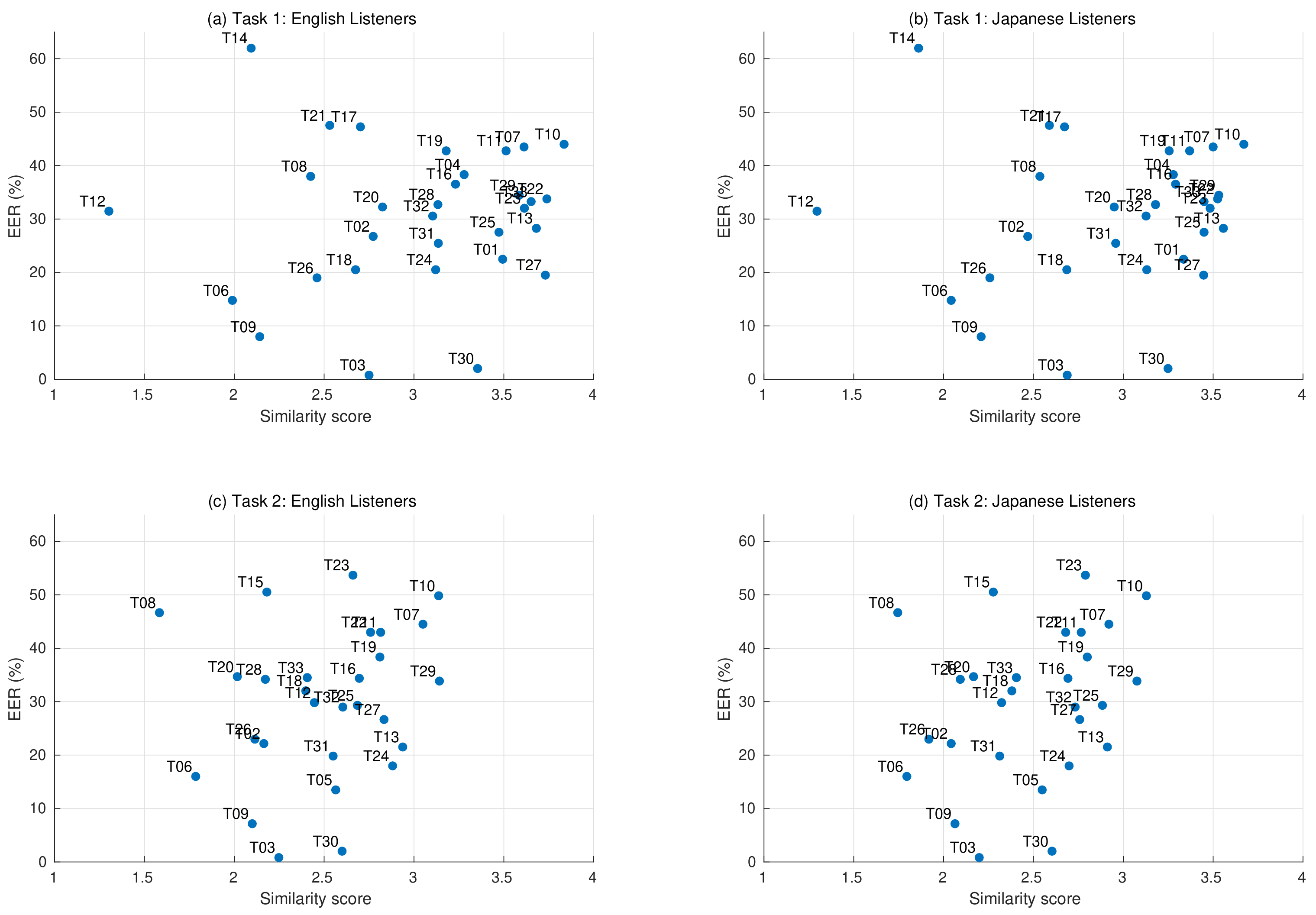}
\end{center}
\caption{Scatter plots for spoofing countermeasure EER (\%) with subjective speaker similarity.} 
\label{corr_CM_EER_SIM}
\end{figure}

\begin{figure}[h!]
\begin{center}
\includegraphics[width=0.85\textwidth]{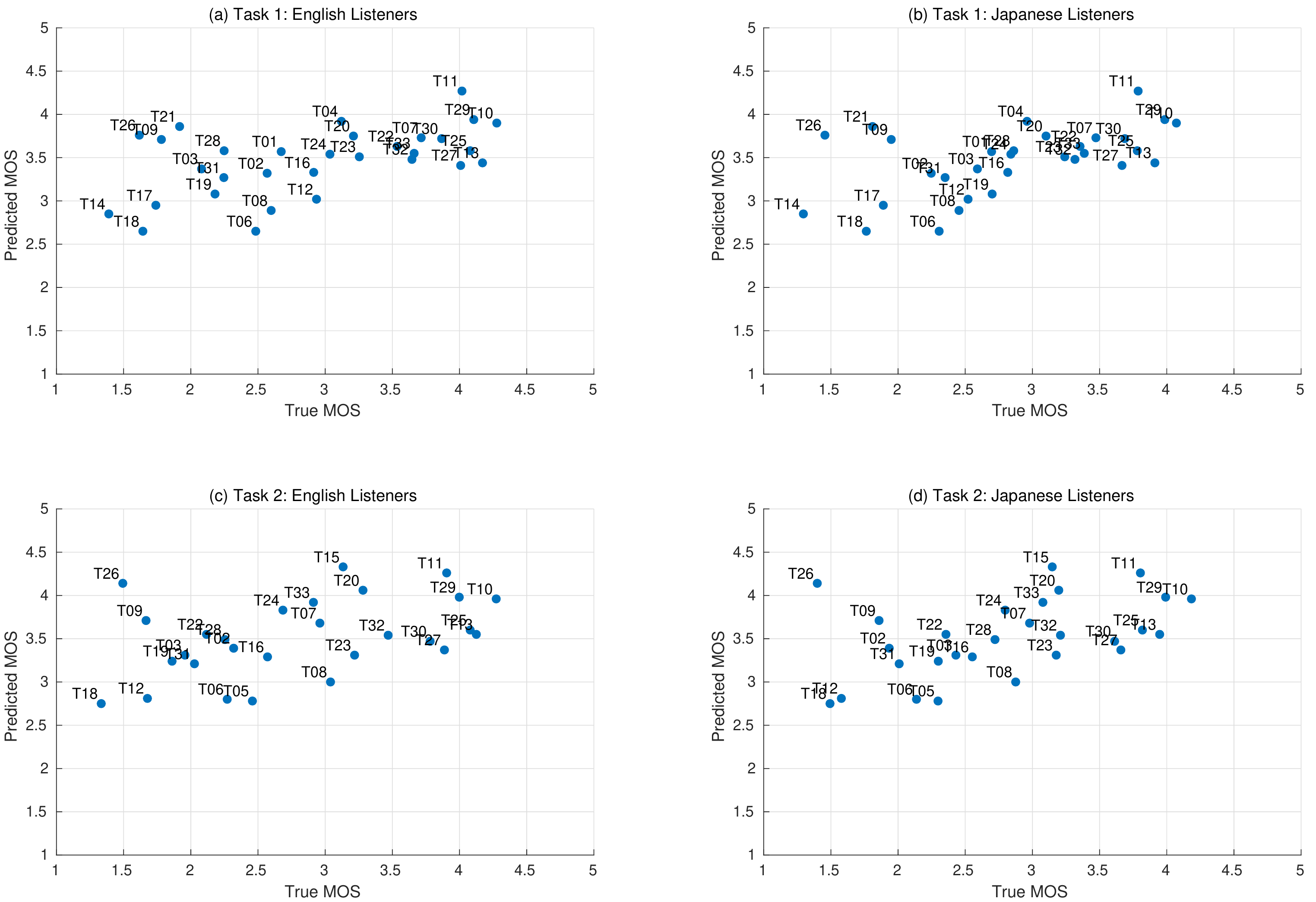}
\end{center}
\caption{Scatter plots for MOSNet (vcc18) predictions with subjective MOS.} 
\label{corr_MOSNet_MOS}
\end{figure}

\begin{figure}[h!]
\begin{center}
\includegraphics[width=0.85\textwidth]{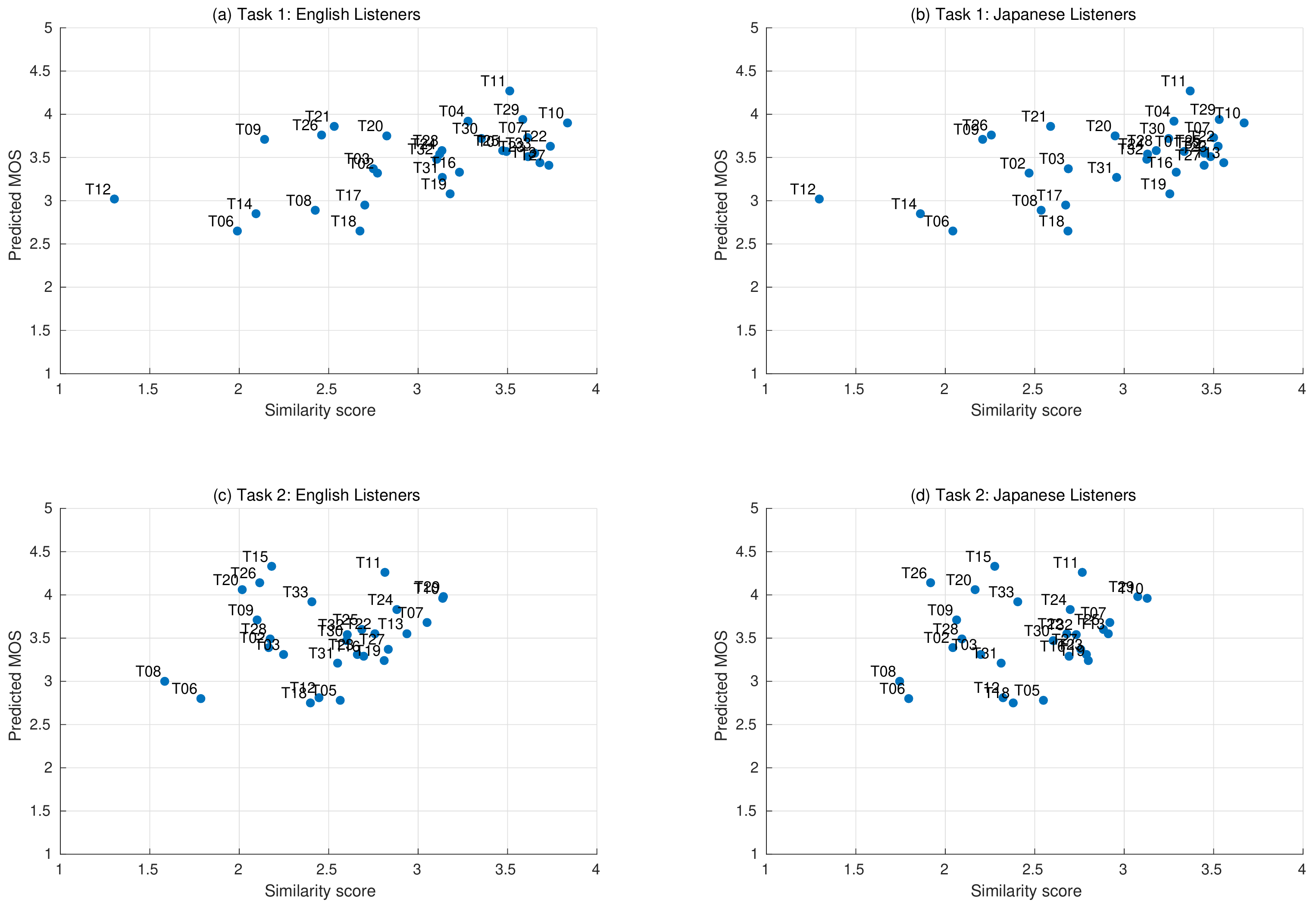}
\end{center}
\caption{Scatter plots for MOSNet (vcc18) predictions with subjective speaker similarity.} 
\label{corr_MOSNet_SIM}
\end{figure}

\begin{figure}[h!]
\begin{center}
\includegraphics[width=0.85\textwidth]{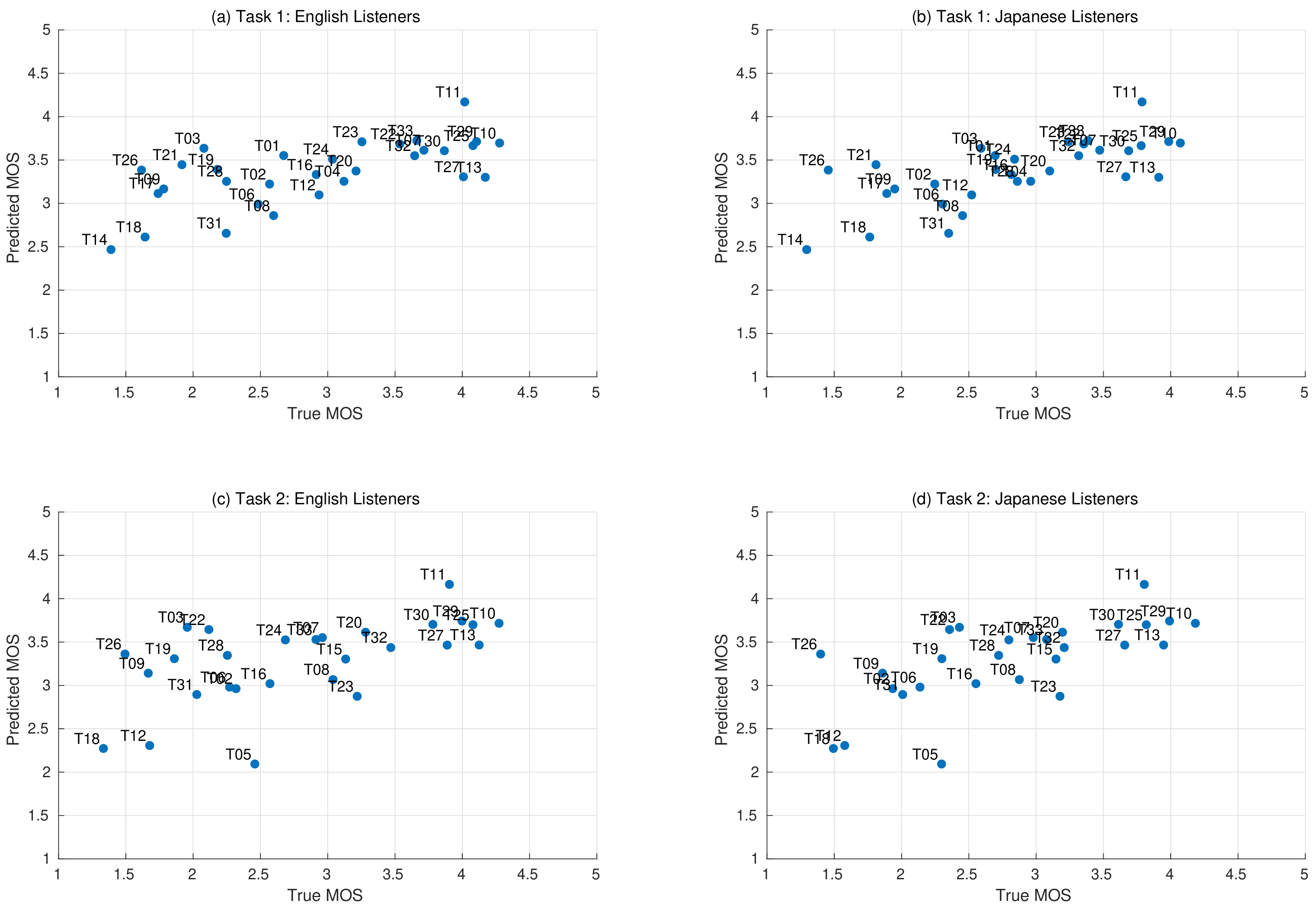}
\end{center}
\caption{Scatter plots for MOSNet (asvspoof19) predictions with subjective MOS.} 
\label{corr_MOSNet19_MOS}
\end{figure}

\begin{figure}[h!]
\begin{center}
\includegraphics[width=0.85\textwidth]{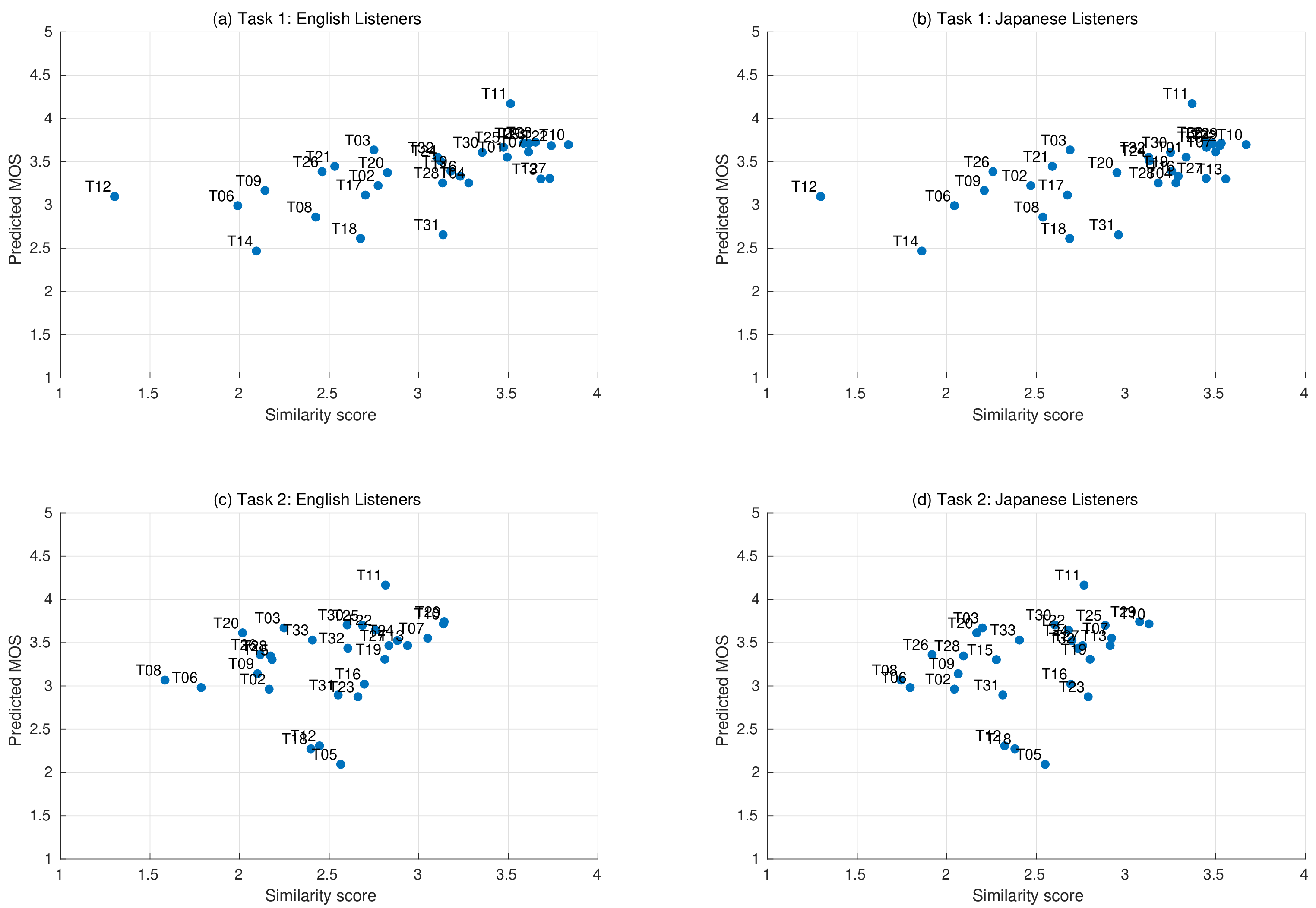}
\end{center}
\caption{Scatter plots for MOSNet (asvspoof19) predictions with subjective speaker similarity.} 
\label{corr_MOSNet19_SIM}
\end{figure}

\begin{figure}[h!]
\begin{center}
\includegraphics[width=0.85\textwidth]{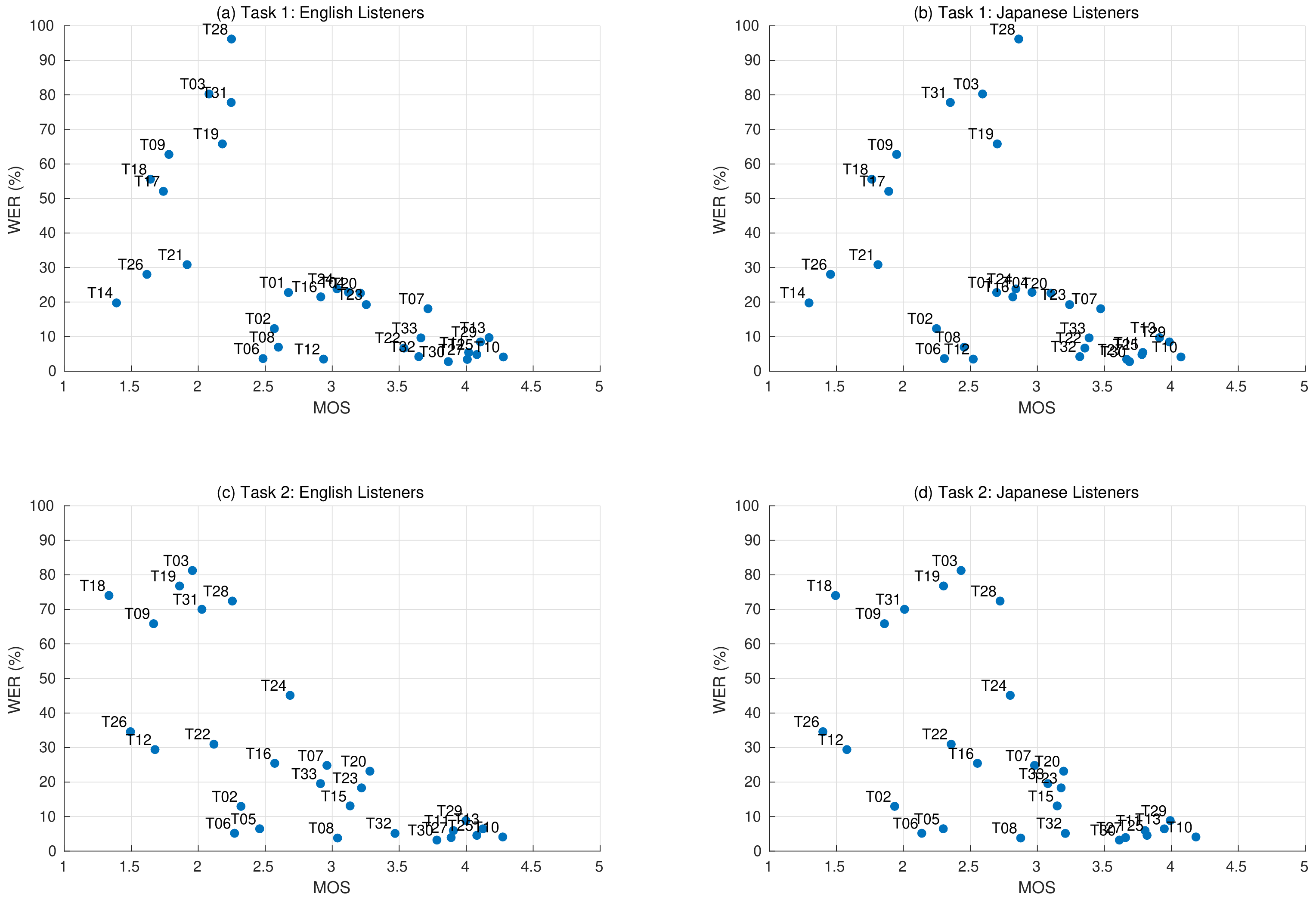}
\end{center}
\caption{Scatter plots for ASR WER (\%) with subjective MOS.} 
\label{corr_ASR_WER_MOS}
\end{figure}

\begin{figure}[h!]
\begin{center}
\includegraphics[width=0.85\textwidth]{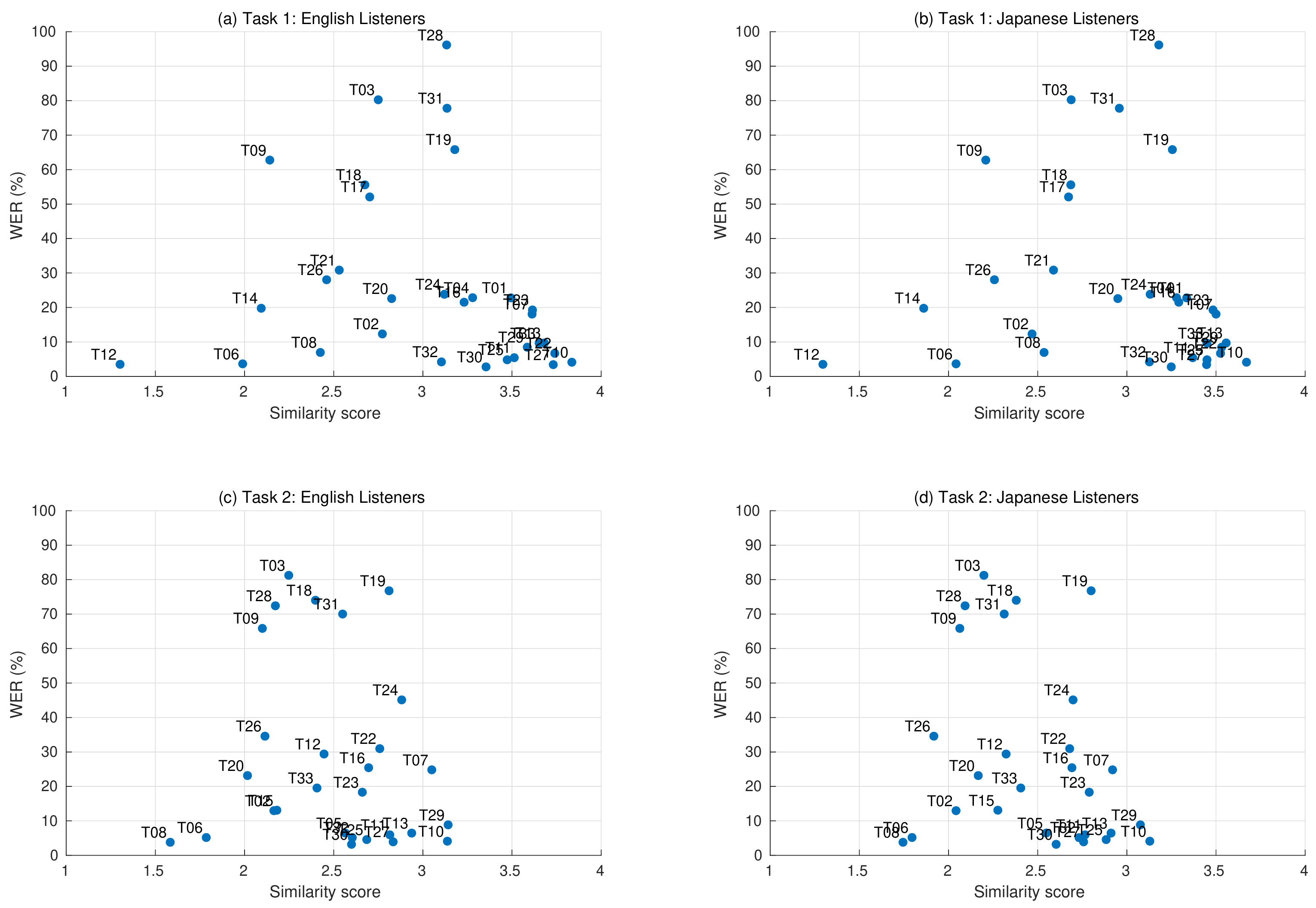}
\end{center}
\caption{Scatter plots for ASR WER (\%) with subjective speaker similarity.} 
\label{corr_AS_WER_SIM}
\end{figure}

The analysis presented here shows the correlation scatter plots of various objective measures against the subjective MOS and speaker similarity for both English and Japanese listeners. The Pearson correlation coefficients along with the $p$-values corresponding to Figures~\ref{corr_ASV_EER_MOS}–~\ref{corr_AS_WER_SIM} have been presented in Table~\ref{tab:correlation}.

\clearpage
\section{Objective Evaluation Results by Target Speaker Language}
\label{appendix:language}

Here, we perform an analysis on the effect of target speaker language on Task 2 of VCC 2020 for different objective measures on all the systems. The correlation of each objective measure with the subjective test for the language pairs is presented in Table~\ref{tab:correlation-breakdown}. In the figures, ``Fin", ``Ger", and ``Man" stand for Finnish, German, and Mandarin, respectively.

\begin{figure}[h!]
\begin{center}
\includegraphics[width=0.63\textwidth]{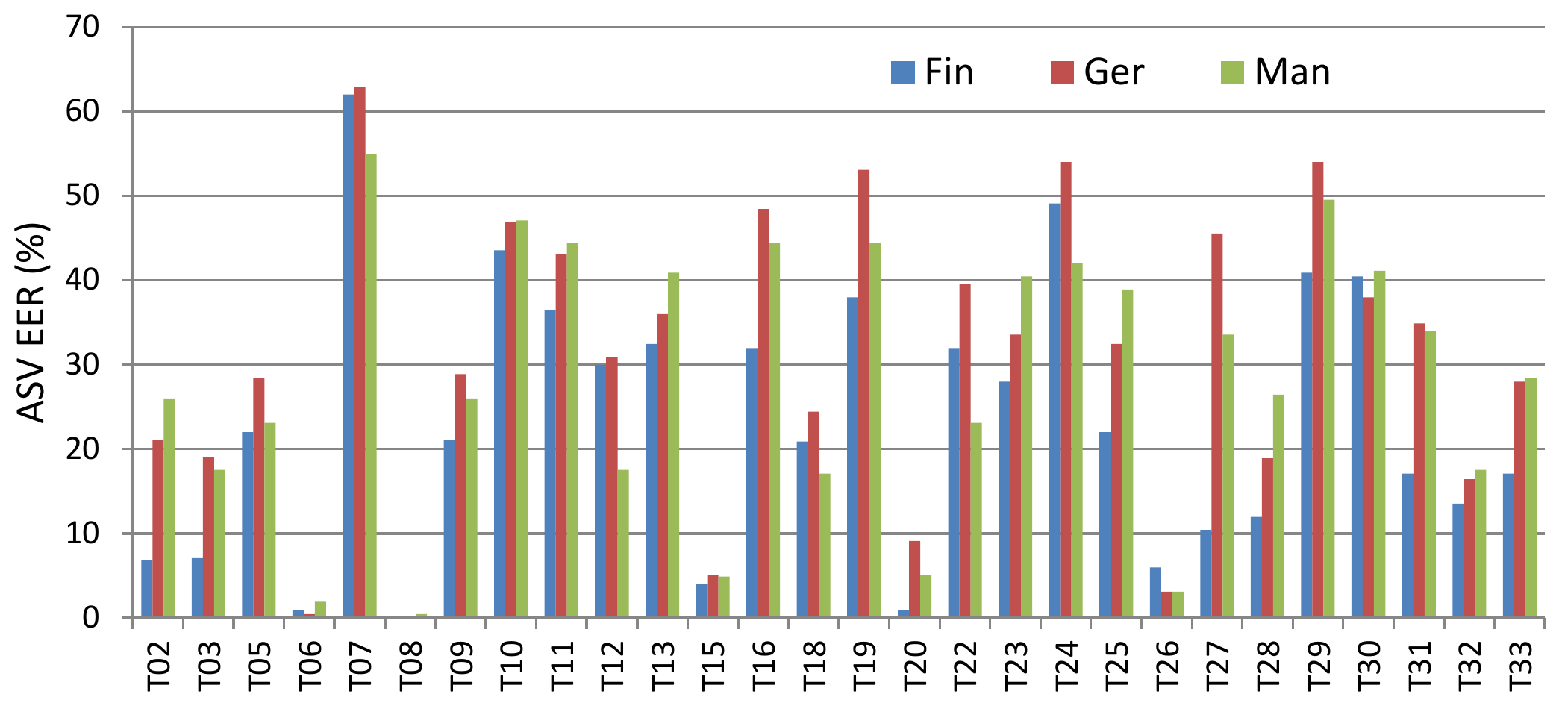}
\end{center}
\caption{ASV performance in EER (\%) of various teams with different language pair analysis in Task 2 of VCC 2020.}
\label{ASVeer_language}
\end{figure}

\begin{figure}[h!]
\begin{center}
\includegraphics[width=0.63\textwidth]{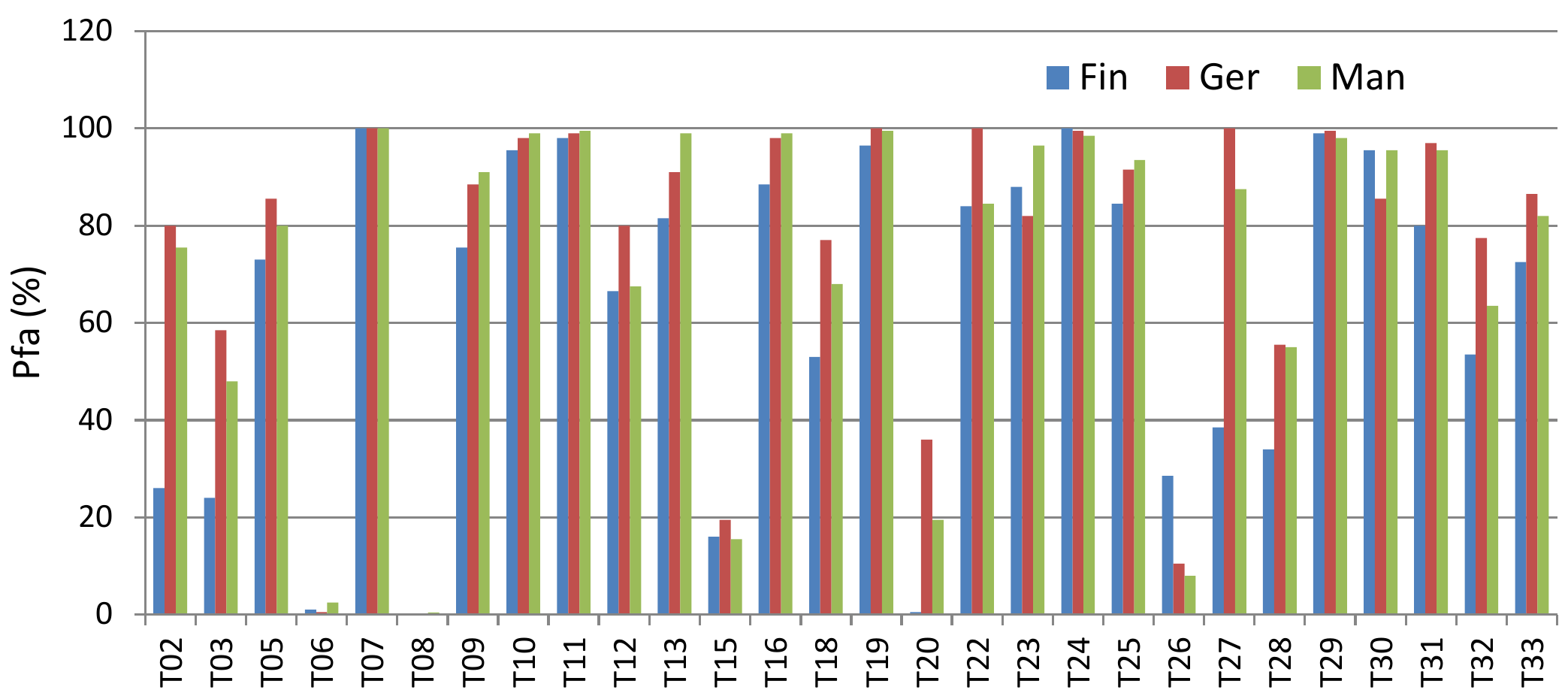}
\end{center}
\caption{ASV performance in Pfa (\%) of various teams with different language pair analysis in Task 2 of VCC 2020.}
\label{ASVpfa_language}
\end{figure}

\begin{figure}[h!]
\begin{center}
\includegraphics[width=0.63\textwidth]{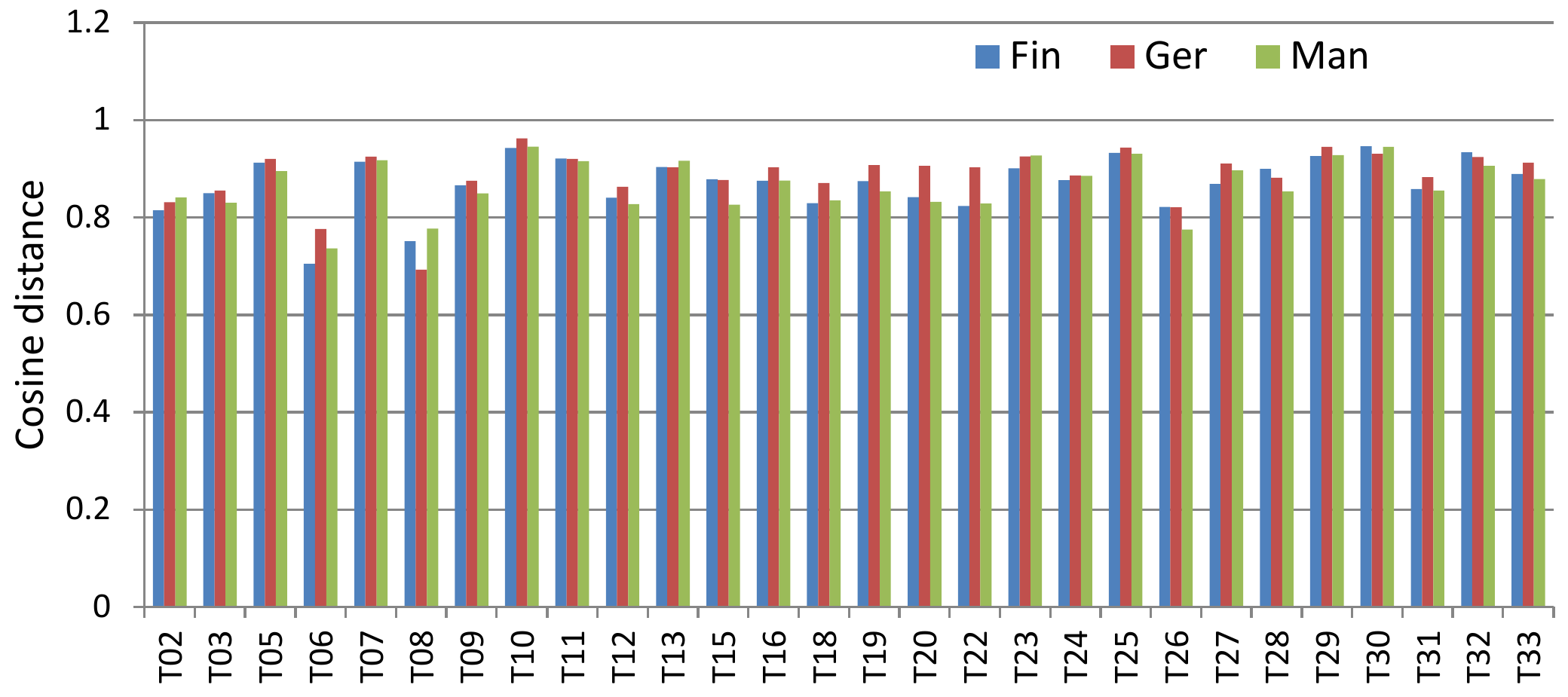}
\end{center}
\caption{Neural speaker embedding cosine similarity of various teams with different language pair analysis in Task 2 of VCC 2020.}
\label{cosine_language}
\end{figure}

\begin{figure}[h!]
\begin{center}
\includegraphics[width=0.63\textwidth]{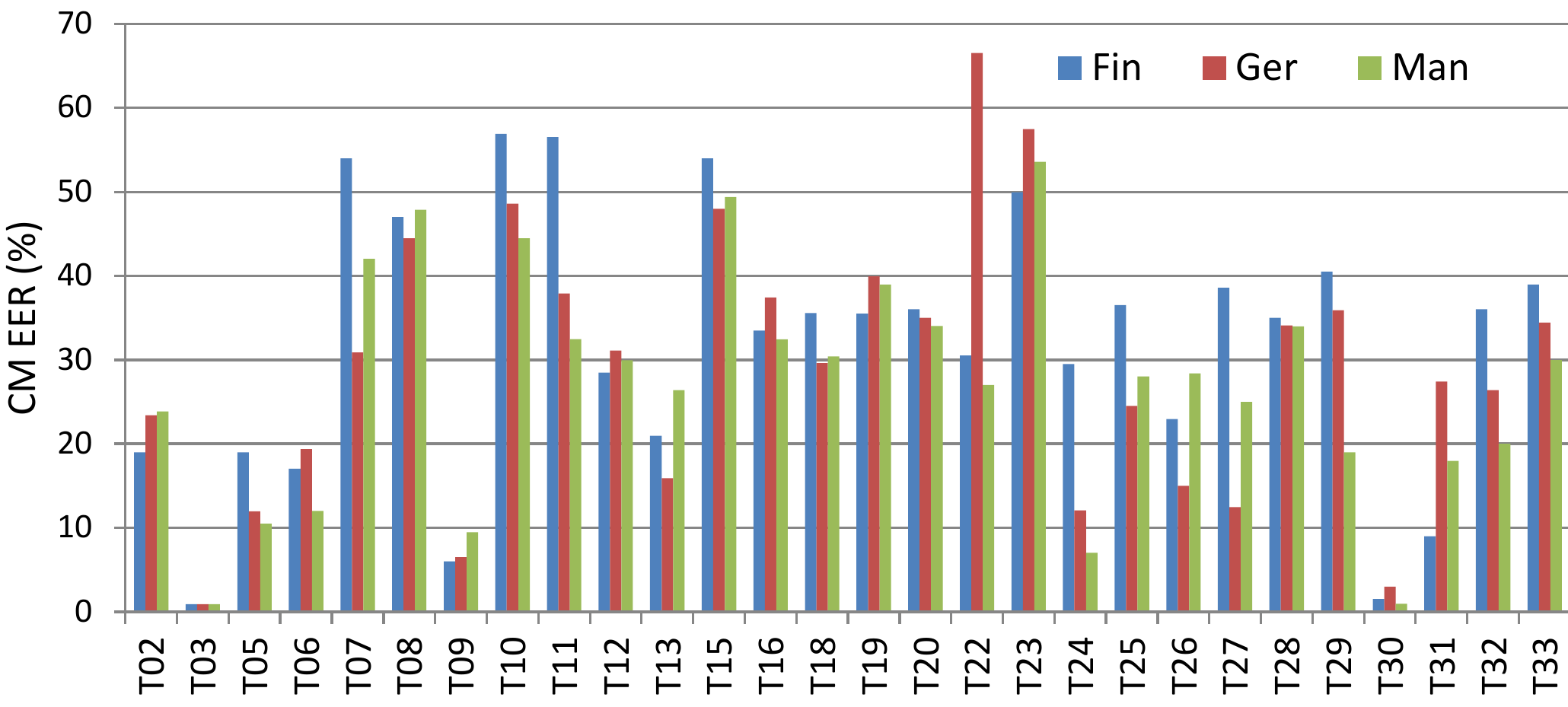}
\end{center}
\vspace{-2mm}
\caption{Spoofing countermeasure performance in EER (\%) of various teams with different language pair analysis in Task 2 of VCC 2020.}
\label{CM_language}
\end{figure}

\begin{figure}[h!]
\begin{center}
\includegraphics[width=0.63\textwidth]{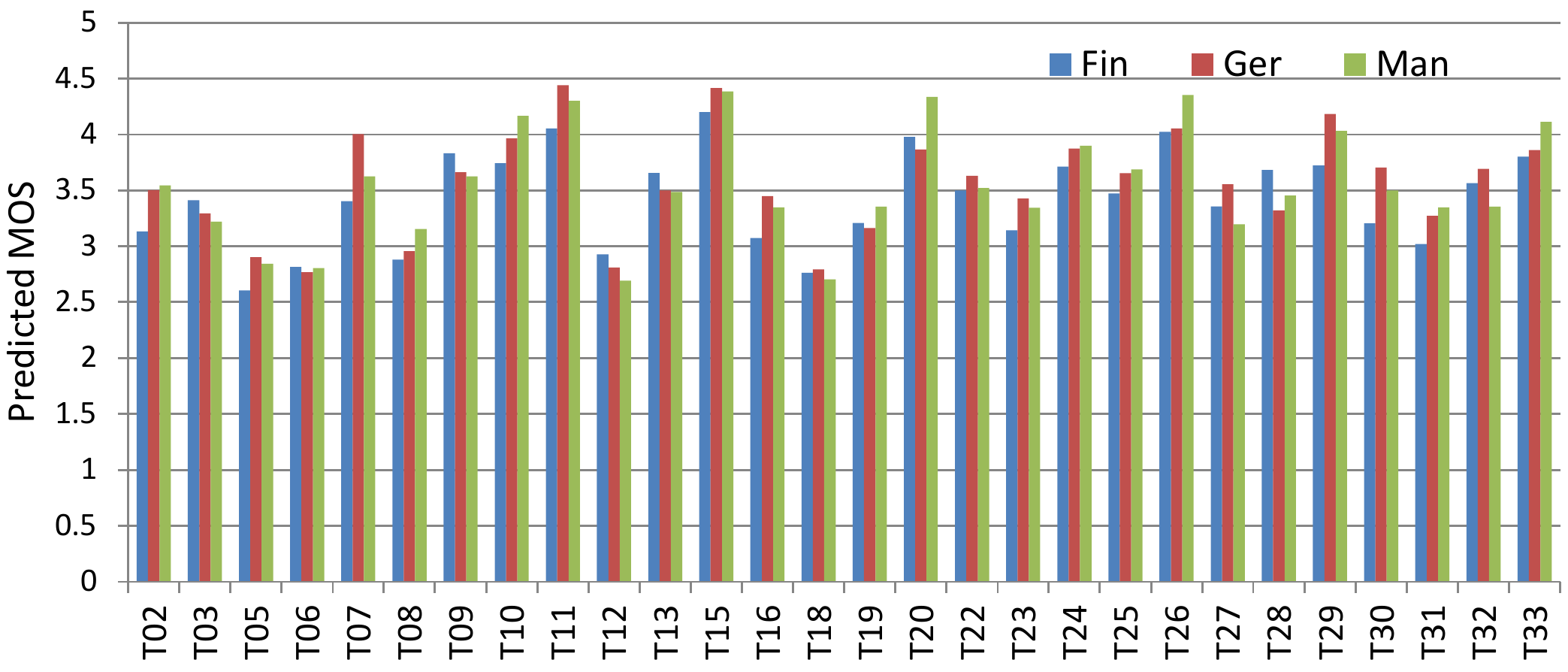}
\end{center}
\vspace{-2mm}
\caption{MOSNet (vcc18) predictions of all systems for different target speaker languages on Task 2 of VCC 2020.}
\label{fig:mos18lang_task2}
\end{figure}

\begin{figure}[h!]
\begin{center}
\includegraphics[width=0.63\textwidth]{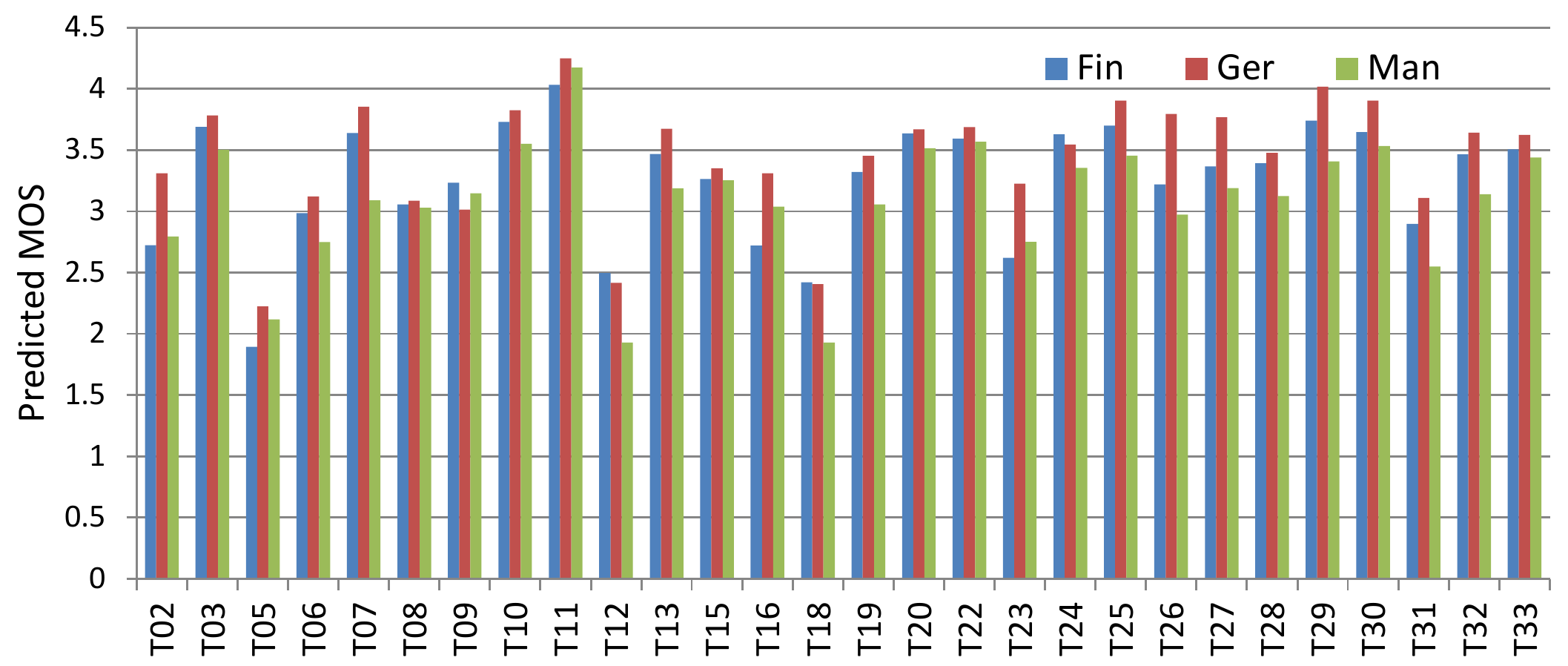}
\end{center}
\vspace{-2mm}
\caption{MOSNet (asvspoof19) predictions of all systems for different target speaker languages on Task 2 of VCC 2020.}
\label{fig:mos19lang_task2}
\end{figure}

\begin{figure}[h!]
\begin{center}
\includegraphics[width=0.63\textwidth]{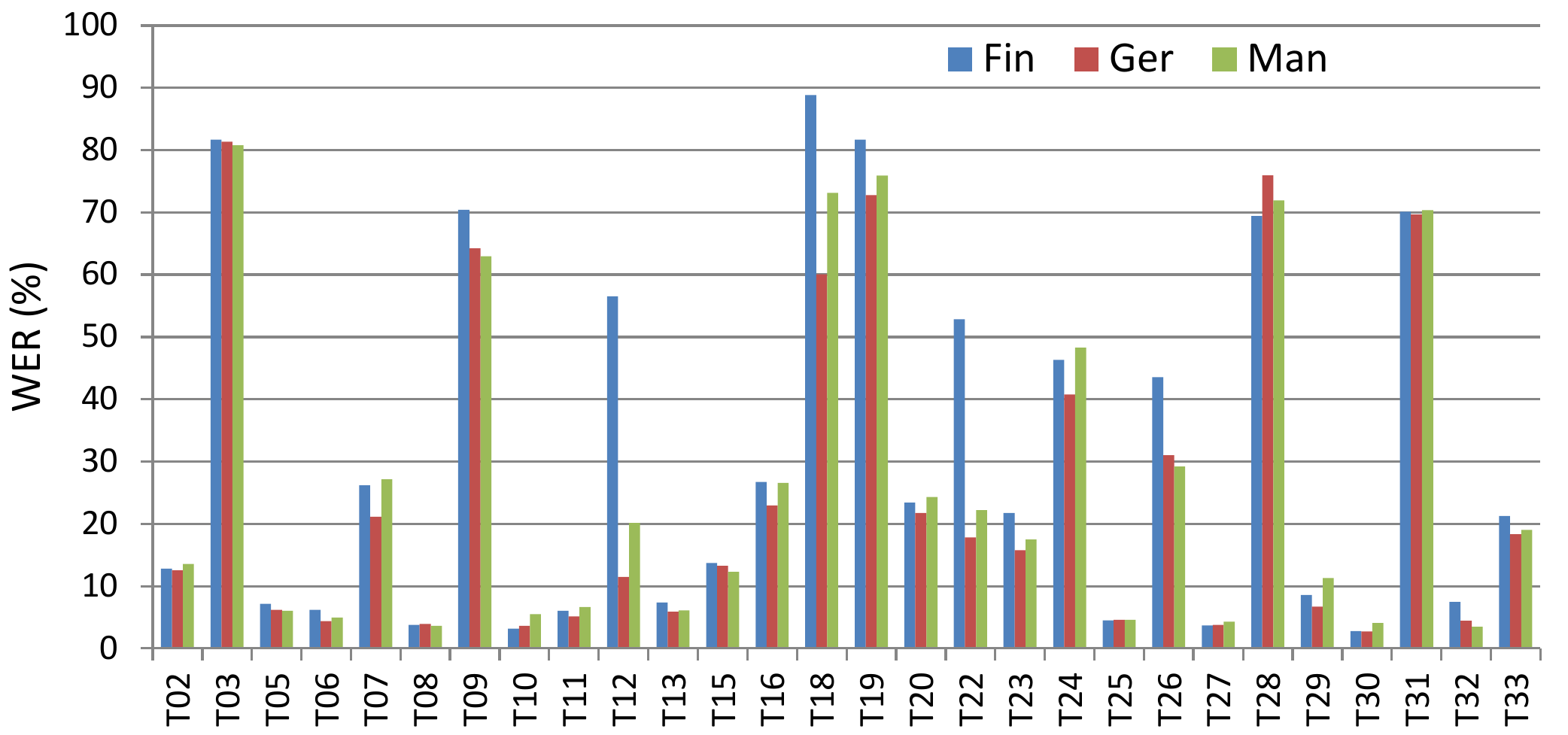}
\end{center}
\vspace{-2mm}
\caption{ASR performance in WER (\%) of various teams with different language pair analysis in Task 2 of VCC 2020.}
\label{fig:ASRwer_lang_task2}
\vspace{-2mm}
\end{figure}

\end{document}